\documentclass[twocolumn,twoside]{IEEEtran}

\ifCLASSINFOpdf

\else

\fi

\ifCLASSOPTIONcompsoc
    \usepackage[caption=false, font=normalsize, labelfont=sf, textfont=sf]{subfig}
\else
    \usepackage[caption=false, font=normalsize]{subfig}
\fi
\usepackage{lipsum}%
\usepackage[dvipsnames]{xcolor}
\usepackage{algorithm,algorithmic}
\usepackage{balance}
\usepackage{multicol}   % for equalize of last page columns1	
\usepackage{cite}
\usepackage{gensymb}
\usepackage{multirow}
\usepackage{graphics}
\usepackage{epsfig}
\usepackage{graphicx}
\usepackage{epstopdf}
\usepackage{textcomp}
\usepackage{amsmath}
\usepackage{mathtools}
\usepackage{filecontents}
\usepackage{lipsum,color}
\usepackage{amssymb}
\usepackage{float}
\usepackage{colortbl} % for painting tables
\usepackage{times} % assumes new font selection scheme installed
\usepackage{amsthm}  % for proof
\usepackage{amsfonts}

\theoremstyle{break}
\begin{document}
\title{A General Model for Pointing Error of High Frequency Directional Antennas}

\author{ Mohammad.~T.~Dabiri,~Mazen~Hasna,~{\it Senior Member,~IEEE},
	~Nizar~Zorba,~{\it Senior Member,~IEEE},
	~Tamer~Khattab,~{\it Senior Member,~IEEE},~and
	~Khalid Qaraqe,~{\it Senior Member,~IEEE}
	%
	%\thanks{This article has been accepted in part for presentation at the IEEE ICC 2022, Seoul, South Korea, May, 2022 \cite{dabiri2022long}.}
	\thanks{Mohammad Taghi Dabiri, Mazen Hasna, Nizar Zorba,  and Tamer Khattab are with the Department of Electrical Engineering, Qatar University, Doha, Qatar.  (E-mail: m.dabiri@qu.edu.qa; hasna@qu.edu.qa; nizarz@qu.edu.qa; tkhattab@ieee.org).}
	\thanks{Khalid A. Qaraqe is with the Department of Electrical and Computer Engineering, Texas A$\&$M University at Qatar, Doha 23874, Qatar (E-mail: khalid.qaraqe@qatar.tamu.edu).}
}

\maketitle
\vspace{-1cm}
%%%%%%%%%%%%%%%%%%%%%%%%%%%%%%%%%%%%%%%%%%%%%%%%%%%%%%%%%%
%%%%%%%%%%%%%%%%%%%%%%%%%%%%%%%%%%%%%%%%%%%%%%%%%%%%%%%%%%
\begin{abstract}
%%%%%%%%%%%%%%%%%%%%%%%%%%%%%%%%%%%%%%%%%%%%%%%%%%%%%%%%%%
%%%%%%%%%%%%%%%%%%%%%%%%%%%%%%%%%%%%%%%%%%%%%%%%%%%%%%%%%%
This paper focuses on providing an analytical framework for the quantification and evaluation of the pointing error for a general case at high-frequency millimeter wave (mmWave) and terahertz (THz) communication links. For this aim, we first derive the the probability density function (PDF) and cumulative distribution functions (CDF) of the pointing error between an unstable transmitter (Tx) and receiver (Rx), that have different antenna patterns and for which the vibrations are not similar in the Yaw and Pitch directions. The special case where the Tx and Rx are both equipped with uniform linear array antenna is also investigated. In addition, using $\alpha-\mu$ distribution, which is a valid model for small-scale fading of mmWave/THz links, the end-to-end PDF and CDF of the considered channel is derived for all the considered cases. Finally, by employing Monte-Carlo simulations, the accuracy of the analytical expressions is verified and the performance of the system is studied.

\end{abstract}
\begin{IEEEkeywords}
Antenna pattern, antenna misalignment, backhaul links, pointing errors, mmWave, THz systems.
\end{IEEEkeywords}
\IEEEpeerreviewmaketitle

%%%%%%%%%%%%%%%%%%%%%%%%%%%%%%%%%%%%%%%%%%%%%%%%%%%%%%%%%%%%
%%%%%%%%%%%%%%%%%%%%%%%%%%%%%%%%%%%%%%%%%%%%%%%%%%%%%%%%%%%%
\section{Introduction}
%%%%%%%%%%%%%%%%%%%%%%%%%%%%%%%%%%%%%%%%%%%%%%%%%%%%%%%%%%%%
%%%%%%%%%%%%%%%%%%%%%%%%%%%%%%%%%%%%%%%%%%%%%%%%%%%%%%%%%%%% https://www.hindawi.com/journals/wcmc/2017/1830987/

\IEEEPARstart{I}{n} the quest to increase network capacity, high-frequency millimeter wave (mmWave) and terahertz (THz) communication links are widely considered to be one of the next frontiers for future wireless systems, because of the wide swaths of unused and unexplored spectrum.
While the mmWave frequencies are already considered in 5G, the vendors and operators have started to look at the usage of the enormous bandwidth of higher THz frequencies \cite{moltchanov2022tutorial} and\cite{petrov2020capacity}.
To mitigate the negative effects of the high path-loss at the mmWave/THz bands, the small wavelength enables the realization of a compact form of highly directive antenna arrays, which also allow a better spatial reuse \cite{chen2022tutorial} and \cite{sarieddeen2021overview}. 
%One of the prospective scenarios is to use the large available bandwidth at mmWave/THz frequencies to provided high data rate point-to-point communication links for aerial and even space nodes such as low- and high-altitude unmanned aerial vehicles (UAVs), flying airplanes, and LEO satellites.
%
This is an important feature for aerial and even space nodes such as low- and high-altitude unmanned aerial vehicles (UAVs), and low earth orbit (LEO) satellites, which are in great needs of large data rates. Therefore, one of the attractive scenarios is to use the large available bandwidth at mmWave/THz frequencies in order to provide extra data rate for point-to-point aerospace communications.

%
%%%%%%%%%%%%%%%%%%%%%%%%%%%%%%%%%%%%%%%%%%%%%%%%%%%%%%%%%%%%%%%%
%%%%%%%%%%%%%%%%%%%%%%%%%%%%%%%%%%%%%%%%%%%%%%%%%%%%%%%%%%%%%%%% VERSUS P_T
\begin{figure}
	\begin{center}
		\includegraphics[width=3.35 in]{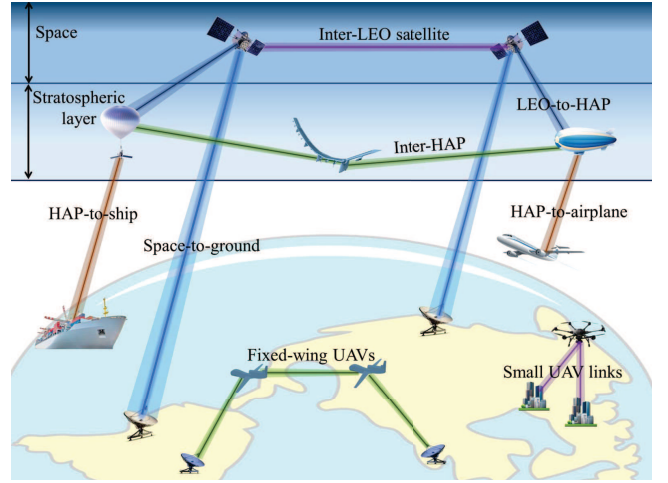}
		\caption{A nominal illustration of the next generation of aerial networks includes a large number of high-frequency and ultra-high data rates between aerial nodes that require the use of directional antennas.}
		\label{sysm}
	\end{center}
\end{figure}
%%%%%%%%%%%%%%%%%%%%%%%%%%%%%%%%%%%%%%%%%%%%%%%%%%%%%%%%%%%%%%%%
%%%%%%%%%%%%%%%%%%%%%%%%%%%%%%%%%%%%%%%%%%%%%%%%%%%%%%%%%%%%%%%%
%
%%%%%%%%%%%%%%%%%%%%%%%%%%%%%%%%%%%%%%%%%%%%%%%%%%%%%%%%%%%%%%%%
%%%%%%%%%%%%%%%%%%%%%%%%%%%%%%%%%%%%%%%%%%%%%%%%%%%%%%%%%%%%%%%% VERSUS W_Z
%\begin{figure*}
%	\centering
%	\subfloat[] {\includegraphics[width=5.35 in]{PP/deviation_in_xy_directions.pdf}
%		\label{n2}
%	}
%	\hfill
%	\subfloat[] {\includegraphics[width=1.4 in]{fig_powerpoint/Yaw_Pitch.pdf}
%		\label{n3}
%	}
%	\caption{Graphical illustration of (b) pointing errors of two directional antennas which are mounted on an unstable Tx and Rx;
%		%
%		(c) the relationship between the axes of the Cartesian coordinate system and the roll, yaw, and pitch directions in the considered system model. }
%	\label{n4}
%\end{figure*}
%%%%%%%%%%%%%%%%%%%%%%%%%%%%%%%%%%%%%%%%%%%%%%%%%%%%%%%%%%%%%%%%
%%%%%%%%%%%%%%%%%%%%%%%%%%%%%%%%%%%%%%%%%%%%%%%%%%%%%%%%%%%%%%%%

One of the main challenges of the tackled scenario is mobility, making the high-directional mmWave/THz antennas suffer from transceivers antenna misalignment, known as pointing error for aerial/mobile nodes.
Therefore, along with other research topics in mmWave/THz bands, studying the effect of pointing errors on system performance is a necessity for establishing a high reliable communication link, which has recently been the subject of several studies \cite{badarneh2022performance,9492775,du2022performance,boulogeorgos2019analytical,tong2021calculating,boulogeorgos2022outage,boulogeorgos2022joint,liu2021thz,LetW}.
In summary, in \cite{badarneh2022performance,9492775,du2022performance,boulogeorgos2019analytical,tong2021calculating,boulogeorgos2022outage,boulogeorgos2022joint,liu2021thz}, the well-known pointing error model provided in \cite{farid2007outage} is used, which, although suitable for optical communication systems and a special case of mmWave/THz systems, as we will show in this work, cannot be directly used for the typical mmWave/THz communication systems.  

More recently, in \cite{LetW} a new pointing error model is provided for mmWave/THz communication links in the presence of an unstable  transmitter (Tx) and receiver (Rx) nodes as a function of real antennas' pattern.
However, the results of \cite{LetW} are obtained for a very special case where the intensity of the vibrations of both  Tx and Rx is the same in the direction of Yaw and Pitch, and also the Tx and Rx nodes have the same antenna pattern with the same stability.
However, as shown in Fig. \ref{sysm}, the Tx and Rx nodes are not necessarily of the same type, not necessarily at the same altitude and atmospheric conditions, and as a result, their instability is not the same. 
As an example, for a communication link between a small multi-rotor UAV and a balloon, it is clear that their instability is different, and more importantly, because the balloon is capable of carrying more payload, it is possible to use a larger antenna for the balloon.
In addition, the intensity of the angular vibrations of the nodes is not necessarily the same in the directions of Yaw and Pitch. 
For example, although for some types of hovering UAVs the intensity of vibrations in both directions of Yaw and Pitch is almost the same, for fixed-wing UAVs or airplanes, the intensity of vibrations in the direction of Pitch is greater than the intensity of vibrations in the direction of Yaw. 
%Therefore, below, for the general case (the Tx and Rx nodes have different antenna patterns and angular intensities, and also have different values for the intensity of vibrations in the directions of Yaw and Pitch), we derive the PDF and CDF of pointing error.
%
%
%
Therefore, the model provided in \cite{LetW} is used only in the special scenario where the Tx and Rx are perfectly symmetric, standing as a major drawback if wanted to use in practical systems. In order to design and analyze the mmWave/THz connections between the moving nodes, it seems necessary to provide a general and tractable model for pointing errors based on the severity of Tx and Rx vibrations, which is the main contribution of this work. %it is necessary to present a more comprehensive model for pointing error.
%As the frequency increases, it becomes more possible to use and support higher link directionality and can be achieved in much smaller footprints and pencil beam, and thus, the system performance becomes more sensitive to pointing errors. Therefore, it seems necessary to provide a general and tractable model for pointing errors based on the severity of Tx and Rx vibrations which is the main contribution of this work. 

In this work, we consider a general case between two unstable Tx and Rx nodes that are completely asymmetric, meaning that they have different antenna patterns and the vibrations of the nodes are not necessarily the same in the directions of Yaw and Pitch. 
For the considered general scenario, we derive the probability density function (PDF) and the cumulative distribution functions (CDF) of pointing error by tacking into account the 3D real antenna pattern. We show that the obtained closed-form models are completely different from the pointing error models of an optical link, and therefore, the pointing error models of the optical channels cannot be directly used to model the pointing error of mmWave/THz links. 
Then, for a special case that the standard deviation (SD) of vibrations are approximately equal in the directions of Yaw and Pitch, we derive a more tractable model for the PDF and CDF of pointing error. Notice that, for the special case where the Tx and Rx nodes are completely symmetric (meaning that the SD of vibrations of Tx and Rx nodes along with the antenna pattern are the same), the pointing error models provided in this work reduce to the simple pointing error model provided in \cite{LetW}. 
In some situations, such as communications between Low-altitude UAVs, the use of a uniform linear array (ULA) antenna in the direction perpendicular to the ground can have several advantages compared to the planar array antenna. 
Therefore, for the case where the Tx and Rx are equipped with ULA antennas, we also derive the PDF and CDF of pointing error.
In addition, using $\alpha-\mu$ distribution, which is a valid model for small-scale fading of higher frequences (i.e., mmWave and THz links), the end-to-end PDF and CDF of the considered channel are derived for all the mentioned cases.
Finally, by employing Monte-Carlo simulations, the accuracy of the analytical expressions is verified and the performance of the system is studied.

The remainder of the paper is structured as follows.
In Section II, the channel of a point-to-point communication link between two unstable Tx and Rx is characterized. The analytical derivations for pointing error are provided in Section III along with several numerical and simulation results to verify the accuracy of the derivations. In Section IV, the analytical derivations for the end-to-end channel are provided for different scenarios. At the end of Section IV, by providing several simulation and analytical results, the performance of the considered system is studied.
Finally, conclusions are drawn in Section V.

%
%%%%%%%%%%%%%%%%%%%%%%%%%%%%%%%%%%%%%%%%%%%%%%%%%%%%%%%%%%%%%%%%
%%%%%%%%%%%%%%%%%%%%%%%%%%%%%%%%%%%%%%%%%%%%%%%%%%%%%%%%%%%%%%%% VERSUS P_T
\begin{figure}
	\begin{center}
		\includegraphics[width=3.4 in]{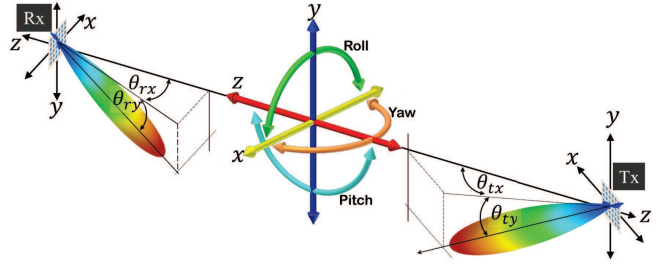}
		\caption{ Graphical illustration of pointing errors of two directional antennas which are mounted on an unstable Tx and Rx.
			The relationship between the axes of the Cartesian coordinate system and the roll, Yaw, and Pitch directions are depicted in the considered system model. }
		\label{n4}
	\end{center}
\end{figure}
%%%%%%%%%%%%%%%%%%%%%%%%%%%%%%%%%%%%%%%%%%%%%%%%%%%%%%%%%%%%%%%%
%%%%%%%%%%%%%%%%%%%%%%%%%%%%%%%%%%%%%%%%%%%%%%%%%%%%%%%%%%%%%%%%
%

%+pencil-beam ++++THz systems can also support higher link directionality and can be achieved in much smaller footprints.

%High bandwidth applications will always benefit from the higher bandwidths offered by mmWave and THz communications with the latter having potential to provide the extreme data rates, but only in low loss situations.

%---------- lens

%
%--------------------------------
%--------------------------------
\section{The System Model}
%--------------------------------
%--------------------------------
Consider an optical receiver where a lens is used to focus the incoming received signal onto the small detector or feed antenna located at the focal point. 
In this case, if the center of received Gaussian beam deviates by $d_v$ from the center of the receiver lens, the fraction of the collected power by the the circular lens with radius $a$ (called the pointing error) is calculated as \cite{farid2007outage,dabiri2018channel}
\vspace{-.2 cm}
\begin{align}
	\label{hp1}
	h_\text{p} = \int_{-a}^a \int_{-\sqrt{a^2-x^2}}^{\sqrt{a^2-x^2}} \frac{2}{\pi w_z^2}
	\exp\left(     -2 \frac{(x-d_v)^2+y^2}{w_z^2}  
	\right)   \text{d}x\text{d}y,
\end{align}
where $w_z$ is the beamwidth of the received Gaussian beam. Using Eq. \eqref{hp1} and after a series of manipulations, the distribution of $h_p$ is derived in \cite[Eq. (11)]{farid2007outage}.
As mentioned earlier, in the pointing error studies of THz communications, it is typical to refer to \cite[Eq. (11)]{farid2007outage} in order to model the pointing error. Although this model is suitable for FSO systems and/or a specific class of high frequency communication links that use a circular lens in the receiver, in this work, we show that this model is not an accurate model for Lens-less mmWave/THz communication systems.

Consider Fig. \ref{n4}, where we tackle a point-to-point high frequency communication system between two unstable nodes, and where Tx and Rx are equipped with high directional antennas. The channel model can be written as \cite{kokkoniemi2020impact,balanis2016antenna}
\begin{align}
	\label{he1}
	h           =   h_L h_a \sqrt{G_t(\theta_t,\phi_t) G_r(\theta_r,\phi_r)},
\end{align}
where $G_q(\theta_q,\phi_q)$ is the antenna radiation pattern defined in Eq. \eqref{f_1}, $h_a$ is the small scale fading, $h_L=h_{Lf}h_{Lm}$ is the channel path loss, $h_{Lf}=\left(\frac{\lambda}{4\pi Z}\right)^2$ is the free-space path loss, and $h_{Lm}= e^{-\frac{\mathcal{K}(f)}{2}Z}$ denotes the molecular absorption loss where $\mathcal{K}(f)$ is the frequency dependent absorption coefficient.
The experimental results show that water vapor dominates the molecular absorption loss at high frequencies \cite{tripathi2021millimeter}.
In mmWave/THz channels, the $\alpha-\mu$ distribution is one of the most widely used models employed for the distribution function of random variables (RV) $h_a$ which is given as \cite{papasotiriou2021experimentally}
\begin{align}
	\label{he3}
	f_{h_a}(h_a) = \frac{\alpha \mu^\mu}{\hat{h_a}^{\alpha \mu} \Gamma(\mu)} h_a^{\alpha \mu -1}  \exp\left( -\mu \frac{ h_a^\alpha}{\hat{h_a}^\alpha}  \right),
\end{align}
where $\Gamma(\cdot)$ is the Gamma function, $\alpha>0$ is a fading parameter, $\mu$ is the normalized variance of the fading channel envelope, and $\hat{h_a}$ is the $\alpha$-root mean value of the fading channel envelope.

In Eq. \eqref{he1}, the parameter $G_q(\theta_q,\phi_q)$ is the antenna radiation pattern in the directions of $\theta_q$ and $\phi_q$ where the subscript $q\in\{t,r\}$ determines the Tx and Rx nodes.  
Here, we consider a standard uniform $N\times N$ array antenna. 
%the amplitude excitation of the entire array is uniform and
%the spacing between antenna elements equal to $d_a = \lambda/2$.
%
%+++++++++++++++++ For a standard linear array (SLA), where the element spacing is $\lambda/2$ ++++ uniformly weighted
%
By taking into account the effect of all elements \cite{balanis2016antenna}, the array radiation gain in the direction of $\theta_q$ and $\phi_q$ will be 
$G_q(\theta_q, \phi_q) = G_0(N_q) G'_q(\theta_q, \phi_q)$, where 
\vspace{-.4 cm}
\begin{align}
	\label{f_1}
	G'_q(N_q,\theta_q, \phi_q) &=
	\left( \frac{\sin\left(\frac{N_q (k d_{x} \sin(\theta_q)\cos(\phi_q))}{2}\right)} 
	{N_q\sin\left(\frac{k d_{x} \sin(\theta_q)\cos(\phi_q)}{2}\right)}
	\right. \nonumber \\
	%------------
	&\times \left. \frac{\sin\left(\frac{N_q (k d_{y} \sin(\theta_q)\sin(\phi_q))}{2}\right)} 
	{N_q\sin\left(\frac{k d_{y} \sin(\theta_q)\sin(\phi_q)}{2}\right)}\right)^2,
\end{align}
and  $d_{x}=d_y=\frac{\lambda}{2}$ are the spacing between the elements along the  $x$ and $y$ axes, respectively,
$k=\frac{2\pi}{\lambda}$ denotes the wave number, $\lambda=\frac{c}{f_c}$ represents the wavelength, $f_c$ denotes the carrier frequency and $c$ is the speed of light. 
%--------
Also, in order to guarantee that the total radiated power of antennas with different $N$ are the same, the coefficient $G_0$ is defined as
\begin{align}
	\label{cv}
	G_0(N_q)=\frac{4\pi}{\int_0^{\pi}\int_0^{2\pi} G'_q(\theta_q,\phi_q) \sin(\theta_q) d\theta_q d\phi_q}.
\end{align}
Based on Eq. \eqref{f_1}, the maximum value of the antenna gain is equal to $G_0(N)$, which is obtained when $\theta_q=0$.

As shown in Fig. \ref{n4}, we assume that the Tx and Rx are located on axis $z$ and at a distance $Z$ from each other, and both the Tx and Rx try to place the main lobe of the antenna pattern on the $z$ axis. The use of mmWave/THz high-gain antennas makes them more sensitive to antenna misalignment or pointing errors, especially for mobile or aerospace communications. 
The angular fluctuations of each mobile node are modeled in three directions; namely: Yaw, roll, and Pitch. Due to the symmetry in the main-lobe of the antenna pattern in the direction of roll (angle $\phi_q$), we can neglect the effect of orientation deviations in the roll direction on the pointing error. Hence,
the orientation fluctuations in the Yaw and Pitch directions cause a pointing error. 
As shown in Fig. \ref{n4}, orientation deviations in the directions of Yaw and Pitch are equivalent to the orientation deviations in $x-z$ and $y-z$ planes, respectively. Let $\theta_{tx}\sim\mathcal{N}(0,\sigma_{\theta_{tx}})$ and $\theta_{ty}\sim\mathcal{N}(0,\sigma_{\theta_{ty}})$ denote the orientation fluctuations of Tx in $x-z$ and $y-z$ planes, respectively, while $\theta_{rx}\sim\mathcal{N}(0,\sigma_{\theta_{rx}})$ and $\theta_{ry}\sim\mathcal{N}(0,\sigma_{\theta_{ry}})$ denote the orientation fluctuations of Rx in $x-z$ and $y-z$ planes, respectively. 
Therefore, in our model, the parameters $\theta_{q}$ can be defined as  functions of RV $\theta_{qx}$ and $\theta_{qy}$ as follows:
\begin{align}
	\label{f_2}
	\theta_q  &= \tan^{-1}\left(\sqrt{\tan^2(\theta_{qx})+\tan^2(\theta_{qy})}\right).
\end{align}

%--------------------------------
%--------------------------------
\section{Pointing Error Modeling}
%--------------------------------
%--------------------------------
%--------------------------------
In this section, we first derive the distribution function of pointing error for the general case and then provide more tractable and precise analytical expressions for some specific cases.
Moreover, the accuracy of the derived analytical expressions is verified by Monte-Carlo simulations.

Let's rewrite Eq. \eqref{he1} as:
\begin{align}
	\label{he4}
	h           =   h_L h_a h_p,
\end{align}
where $h_p= G_0(N_t,N_r) h'_p$ is the pointing error coefficient, $G_0(N_t,N_r)=\sqrt{G_0(N_t)G_0(N_r)}$ is the maximum antenna gain, and $0<h'_p<1$ is the normalized pointing error
\begin{align}
	\label{he5}
	h'_p          = h'_{pt} h'_{pr} =
	\sqrt{G'_t(N_t,\theta_{tx},\theta_{ty})}
	\sqrt{G'_r(N_t,\theta_{rx},\theta_{ry})}.
\end{align}
From Eq. \eqref{he5}, $h'_p$ is a function of $N_t$ and $N_r$ as well as four RVs $\theta_{tx},\theta_{ty}$, $\theta_{rx}$, and $\theta_{ry}$.

Let us define parameters $\beta_{qw} = \frac{\sigma_{\theta_{qw}}^2}{w_{B_q}}$ where the subscript $w\in\{x,y\}$ determines the vibrations of Yaw (or $x-z$ plane) and Pitch (or $y-z$ plane), and the subscript $q\in\{t,r\}$ determines the Tx and Rx nodes. Based on this definition, the general scenarios refer to the condition that $\sigma_{\theta_{qw}}$s and $w_{Bq}$s (or $N_q$s) have different values.

%%%%%%%%%%%%%%%%%%%%%%%%%%% begin Lemma 2 %%%%%%%%%%%%%%%%%%%%%%%%%%%%%%%%%%%%%%%%%%%%%%%%%%%%%%%%%%%%%%%%%%%%%%%%%%%%%%%%%%%%%%%%%%%%%%%%%%%%%%%%%%%%%%%%%%%%%%%%%%
%%%%%%%%%%%%%%%%%%%%%%%%%%% begin Lemma 2 %%%%%%%%%%%%%%%%%%%%%%%%%%%%%%%%%%%%%%%%%%%%%%%%%%%%%%%%%%%%%%%%%%%%%%%%%%%%%%%%%%%%%%%%%%%%%%%%%%%%%%%%%%%%%%%%%%%%%%%%%%
%%%%%%%%%%%%%%%%%%%%%%%%%%% begin Lemma 2 %%%%%%%%%%%%%%%%%%%%%%%%%%%%%%%%%%%%%%%%%%%%%%%%%%%%%%%%%%%%%%%%%%%%%%%%%%%%%%%%%%%%%%%%%%%%%%%%%%%%%%%%%%%%%%%%%%%%%%%%%%
{\bf Theorem 1.}
{\it  Under the general condition when the parameters $\sigma_{\theta_{qw}}$s and $w_{Bq}$s have different values, the PDF of pointing error is obtained as:}
\begin{align}
	\label{gen1}
	f_{h_p}(h_p) = C_g \frac{h_p^{1/\beta_1-1}}{G_0^{1/\beta_1}}   \sum_{k=0}^K \frac{\Delta_k      \left( - \ln\left(\frac{h_p}{G_0}\right) \right)^{k+1}   }
	{\Gamma(k+2)\beta_1^{k+2}}        	
\end{align}
{\it while its corresponding CDF results in}
\begin{align}
	\label{gen_cdf_1}
	&F_{h_p}(h_p) =1 - \Theta_0
	+ C_g           \left(\frac{h_p}{G_0}\right)^{1/\beta_1}  \nonumber \\
	&~~~\times\sum_{k=0}^K \sum_{j=0}^{k+1} \frac{\Delta_k } {\Gamma(k+2-j)\beta_1^{k+1-j}}    \left(-\ln\left(\frac{h_p}{G_0}\right)\right)^{k+1-j} 
\end{align}
{\it where}
\begin{align}
	\left\{ \!\!\!\!\! \! \!
	\begin{array}{rl}
		&C_g = \prod_{i=1}^4 \sqrt{\beta_1/\beta_i}, \\ 
		&\Delta_k = \frac{1}{k} \sum_{i=1}^k i \gamma_i \Delta_{k+1-i}~~\text{for}~~k=1,...,K, \\
		&\Delta_0=1,~~~
		\gamma_k = \sum_{i=1}^4 \frac{\left(1-{\beta_1}/{\beta_i}\right)^k}{2k}.
	\end{array} \right. \nonumber
\end{align}
{\it Also, $\bar{\beta}=[\beta_1,...,\beta_4]$ where $\beta_1$ is the least value of $\beta_{q_w}$s, $\beta_4$ is the maximum value of $\beta_{q_w}$s, and 
	$\beta_{qw} = \frac{\sigma^2_{\theta_{q_w}}}{w^2_{Bq}}$.}
%%%%%%%%%%%%%%%%%%%%%%%%%%% begin PROOF j_u J_u m_u M_u N_{uqw} \mu_{uqx} \mu_{uqy} \sigma_{uqx}  \sigma_{uqy} %%%%%%%%%%%%%%%%%%%%%%%%%%%%%%%%%%%%%%%%%%%%%%%%%%%%%%%%%%%%%%%%%%%%%%%%%%%%%%%%%%%%%%%%%%%%%%%%%%%%%%%%%%%%%%%%%%%%%%%%%%
%%%%%%%%%%%%%%%%%%%%%%%%%%% begin PROOF %%%%%%%%%%%%%%%%%%%%%%%%%%%%%%%%%%%%%%%%%%%%%%%%%%%%%%%%%%%%%%%%%%%%%%%%%%%%%%%%%%%%%%%%%%%%%%%%%%%%%%%%%%%%%%%%%%%%%%%%%%
\begin{IEEEproof}
	Please refer to Appendix \ref{AppA}. 
\end{IEEEproof}

\begin{figure}
	\centering
	\subfloat[] {\includegraphics[width=3.0 in]{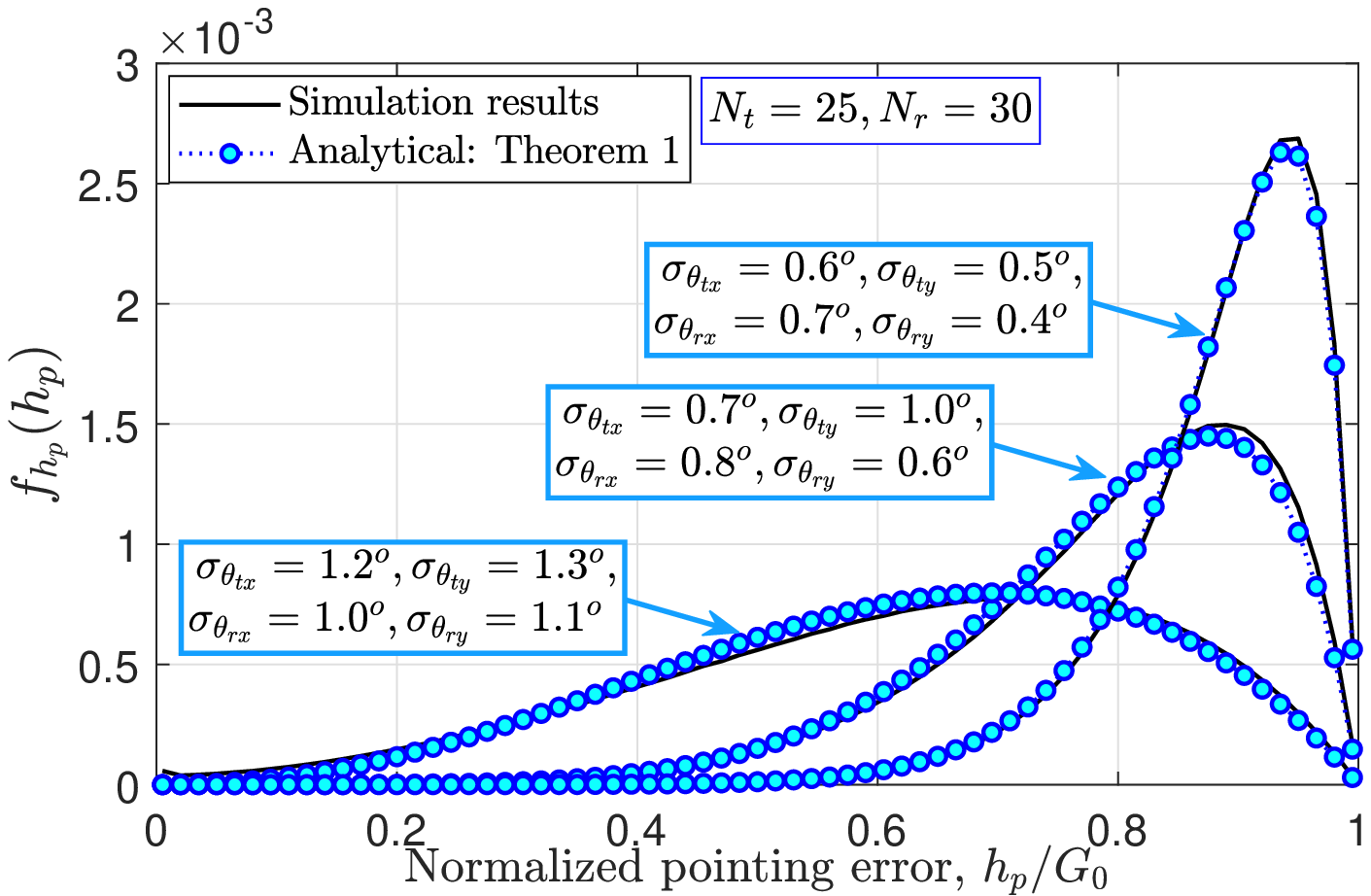}
		\label{gt1}
	}
	\hfill
	\subfloat[] {\includegraphics[width=3.0 in]{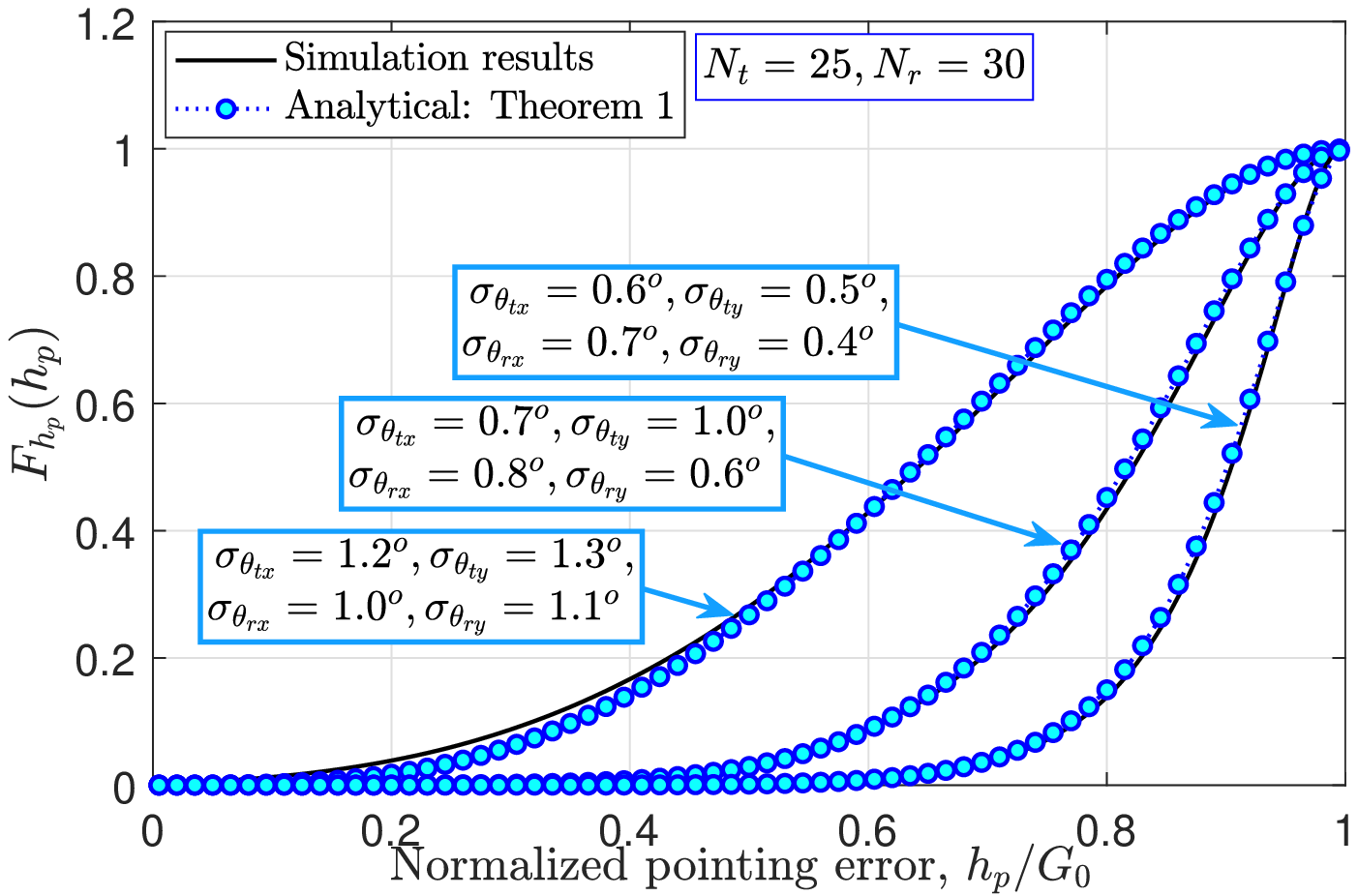}
		\label{gt2}
	}
	\caption{Comparison of the accuracy of the provided analytical results in Theorem 1 with Monte-Carlo simulations for three different states of the intensity of transceiver vibrations (a) for the PDF, and (b) for the CDF.} \label{gt3}
\end{figure}

In Fig. \ref{gt3}, by employing Monte-Carlo simulations, we compare the accuracy of the proposed analytical derivations in Theorem 1 for the three different states of transceiver instability. As mentioned, the vibrations of Tx are characterized by $\sigma_{\theta_{tx}}$ and $\sigma_{\theta_{ty}}$, while the vibrations of  Rx are characterized by $\sigma_{\theta_{rx}}$ and $\sigma_{\theta_{ry}}$.
In the simulations, we use the standard array antenna pattern introduced in Eqs. \eqref{f_1} and \eqref{cv}. To do this, we first generate $5\times10^7$ independent RVs $\theta_{tx}$, $\theta_{ty}$, $\theta_{rx}$, and $\theta_{ry}$, and then, for each $5\times10^7$ independent run, using Eqs. \eqref{f_1}, \eqref{cv}, and \eqref{f_2}, we generate $5\times10^7$ independent RVs of $h_p$. Next, using the generated instantaneous coefficients $h_p$s, the PDF and CDF of pointing error are obtained.
As can be seen, the Monte-Carlo simulations verify the accuracy of the analytical expressions provided in Theorem 1, especially for low $\sigma_{\theta_{qw}}$ values. 
%Note that the provided analytical results are based on the main lobe of the antenna pattern and are only suitable for point-to-point communication.
%
The higher the value of $\sigma_{\theta_{qw}}$, the probability of deviation from the main lobe to the side lobes increases, resulting in a gap between the simulation results and the analytical results. 
For example, for $N_t=25$ and $N_r=30$, although the analytical results for $\sigma_{\theta_{tx}}=0.6^o, \sigma_{\theta_{ty}}=0.5^o, \sigma_{\theta_{rx}}=0.7^o$, and $\sigma_{\theta_{ry}}=0.4^o$ completely match the simulations, by increasing $\sigma_{\theta_{qw}}$s to $\sigma_{\theta_{tx}}=1.2^o, \sigma_{\theta_{ty}}=1.3^o, \sigma_{\theta_{rx}}=1.0^o$, and  $\sigma_{\theta_{ry}}=1.1^o$, it is observed that the analytical results deviate slightly from the simulations.

%%%%%%%%%%%%%%%%%%%%%%%%%%%%%%%%%%%%%%%%%%%%%%%%%%%%%%%%%%%%%%%%
%%%%%%%%%%%%%%%%%%%%%%%%%%%%%%%%%%%%%%%%%%%%%%%%%%%%%%%%%%%%%%%% VERSUS W_Z
%%%%%%%%%%%%%%%%%%%%%%%%%%%%%%%%%%%%%%%%%%%%%%%%%%%%%%%%%%%%%%%%
%%%%%%%%%%%%%%%%%%%%%%%%%%%%%%%%%%%%%%%%%%%%%%%%%%%%%%%%%%%%%%%%

%Note that the results of Theories 1 and 2 are obtained for the planner array antenna.  In some situations, such as communications between Low-altitude UAVs, the use of a linear array antenna in the direction perpendicular to the ground  can have the several advantages compared to the planar array antenna.

%%%%%%%%%%%%%%%%%%%%%%%%%%% begin Lemma 2 %%%%%%%%%%%%%%%%%%%%%%%%%%%%%%%%%%%%%%%%%%%%%%%%%%%%%%%%%%%%%%%%%%%%%%%%%%%%%%%%%%%%%%%%%%%%%%%%%%%%%%%%%%%%%%%%%%%%%%%%%%
%%%%%%%%%%%%%%%%%%%%%%%%%%% begin Lemma 2 %%%%%%%%%%%%%%%%%%%%%%%%%%%%%%%%%%%%%%%%%%%%%%%%%%%%%%%%%%%%%%%%%%%%%%%%%%%%%%%%%%%%%%%%%%%%%%%%%%%%%%%%%%%%%%%%%%%%%%%%%%
%%%%%%%%%%%%%%%%%%%%%%%%%%% begin Lemma 2 %%%%%%%%%%%%%%%%%%%%%%%%%%%%%%%%%%%%%%%%%%%%%%%%%%%%%%%%%%%%%%%%%%%%%%%%%%%%%%%%%%%%%%%%%%%%%%%%%%%%%%%%%%%%%%%%%%%%%%%%%%
{\bf Theorem 2.}
{\it  When $\beta_{tx}=\beta_{ty}=\beta_t$ and $\beta_{rx}=\beta_{ry}=\beta_r$, the PDF and CDF of pointing error are obtained as:}
\begin{align}
	\label{ge2}
	f_{h_p}(h_p) = \frac{ 1}  {G_0(\beta_t-\beta_r)}       
	\left[    \left(\frac{h_p}{G_0} \right)^{1/\beta_t-1}  
	-         \left(\frac{h_p}{G_0} \right)^{1/\beta_r-1} 
	\right],      	
\end{align}
\begin{align}
	\label{ge3}
	F_{h_p}(h_p) =  \frac{ 1 }  {\beta_t-\beta_r} 
	\left[   \beta_t \left(\frac{h_p}{G_0}\right)^{1/\beta_t} 
	-   \beta_r \left(\frac{h_p}{G_0}\right)^{1/\beta_r}   
	\right].
\end{align}

%%%%%%%%%%%%%%%%%%%%%%%%%%% begin PROOF j_u J_u m_u M_u N_{uqw} \mu_{uqx} \mu_{uqy} \sigma_{uqx}  \sigma_{uqy} %%%%%%%%%%%%%%%%%%%%%%%%%%%%%%%%%%%%%%%%%%%%%%%%%%%%%%%%%%%%%%%%%%%%%%%%%%%%%%%%%%%%%%%%%%%%%%%%%%%%%%%%%%%%%%%%%%%%%%%%%%
%%%%%%%%%%%%%%%%%%%%%%%%%%% begin PROOF %%%%%%%%%%%%%%%%%%%%%%%%%%%%%%%%%%%%%%%%%%%%%%%%%%%%%%%%%%%%%%%%%%%%%%%%%%%%%%%%%%%%%%%%%%%%%%%%%%%%%%%%%%%%%%%%%%%%%%%%%%
\begin{IEEEproof}
	Please refer to Appendix \ref{AppB}. 
\end{IEEEproof}

\begin{figure}
	\centering
	\subfloat[] {\includegraphics[width=3.0 in]{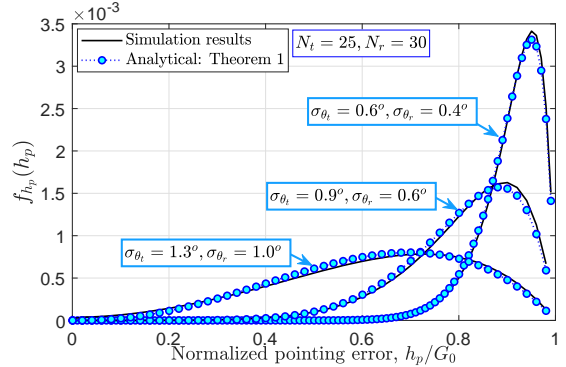}
		\label{nt1}
	}
	\hfill
	\subfloat[] {\includegraphics[width=3.0 in]{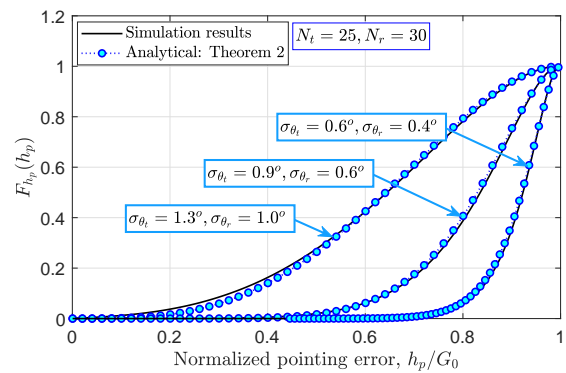}
		\label{nt2}
	}
	\caption{Comparison of the accuracy of the provided analytical results in Theorem 2 with Monte-Carlo simulations for three different states of the intensity of transceiver vibrations (a) for the PDF, and (b) for the CDF.}
	\label{nt3}
\end{figure}

The results of Theorem 2 are for a particular case where the severity of vibrations along the Yaw and Pitch directions are approximately the same or $\sigma_{\theta_{tx}}\simeq\sigma_{\theta_{ty}}=\sigma_\theta$ and $\sigma_{\theta_{rx}}\simeq\sigma_{\theta_{ry}}=\sigma_\theta$. 
For example, for communication between UAVs, when both UAVs are of the same model and at the same altitude, they are expected to have similar fluctuation intensity.
As we observe from Theorem 2, the closed-form expressions provided in Eqs. \eqref{ge2} and \eqref{ge3} are very tractable and simple. In addition, the simulation results of Fig. \ref{nt3} verify the accuracy of the analytical expressions under the different values $\sigma_{\theta_{qw}}$s.

%%%%%%%%%%%%%%%%%%%%%%%%%%%%%%%%%%%%%%%%%%%%%%%%%%%%%%%%%%%%%%%%
%%%%%%%%%%%%%%%%%%%%%%%%%%%%%%%%%%%%%%%%%%%%%%%%%%%%%%%%%%%%%%%% VERSUS W_Z
%%%%%%%%%%%%%%%%%%%%%%%%%%%%%%%%%%%%%%%%%%%%%%%%%%%%%%%%%%%%%%%%
%%%%%%%%%%%%%%%%%%%%%%%%%%%%%%%%%%%%%%%%%%%%%%%%%%%%%%%%%%%%%%%%

\begin{figure}
	\centering
	\subfloat[] {\includegraphics[width=3.0 in]{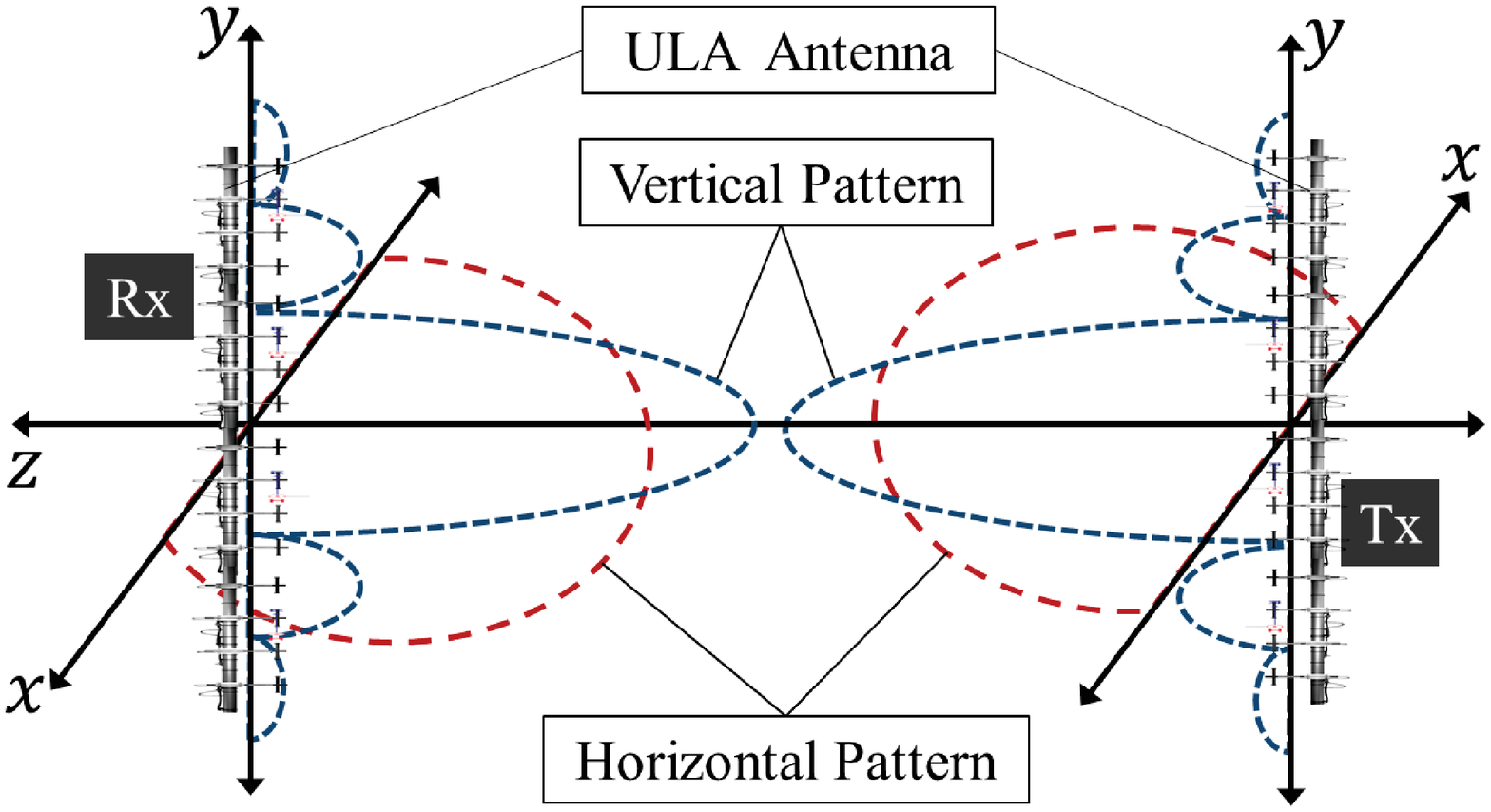}
		\label{ul1}
	}
	\hfill
	\subfloat[] {\includegraphics[width=3.0 in]{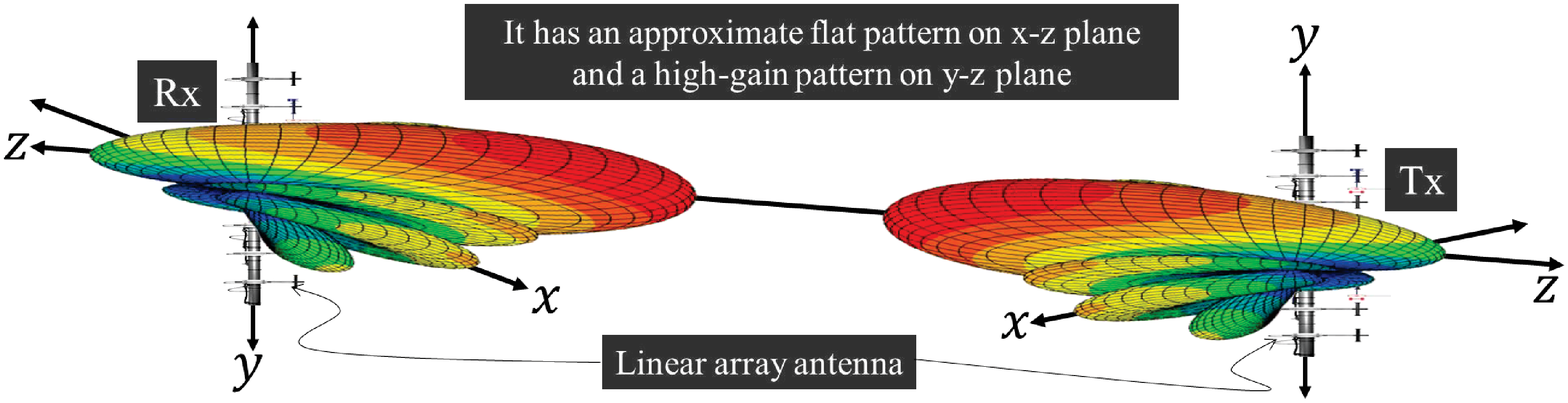}
		\label{ul2}
	}
	\caption{Illustration of considered configuration based on ULA antenna in such a way that the horizontal pattern, which is almost flat, is horizontal to the ground, and the vertical pattern of the antenna is perpendicular to the ground devices:  (a) a better illustration of vertical and horizontal patterns, and (b) a real 3D antenna pattern.}
	\label{ul3}
\end{figure}

The results of Theorems 1 and 2 are for the planner array antenna.
In some situations, such as communications between Low-altitude UAVs, the use of a linear array antenna in the direction perpendicular to the ground (as illustrated in Fig. \ref{ul3}) can have several advantages compared to the planar array antenna. 
The first important feature of using linear area antenna is that the considered communication link will be resistant to vibrations in the direction of the Yaw or x-z plane. This topology has an approximately flat pattern in the x-z plane and therefore, the vibrations of the UAVs in this direction do not have a significant effect on the system performance.
Moreover, this topology has a high-gain pattern in the direction of the Pitch or y-z plane, and thus,  interference caused by ground transmitters is reduced significantly, and vice versa.
In addition, from an aerodynamic point of view, it is a better option for mounting on UAVs than the planar antennas, especially for fixed-wing UAVs that are in motion.

%%%%%%%%%%%%%%%%%%%%%%%%%%%%%%%%%%%%%%%%%%%%%%%%%%%%%%%%%%%%%%%%
%%%%%%%%%%%%%%%%%%%%%%%%%%%%%%%%%%%%%%%%%%%%%%%%%%%%%%%%%%%%%%%% VERSUS W_Z
%%%%%%%%%%%%%%%%%%%%%%%%%%%%%%%%%%%%%%%%%%%%%%%%%%%%%%%%%%%%%%%%
%%%%%%%%%%%%%%%%%%%%%%%%%%%%%%%%%%%%%%%%%%%%%%%%%%%%%%%%%%%%%%%%

%%%%%%%%%%%%%%%%%%%%%%%%%%% begin Lemma 2 %%%%%%%%%%%%%%%%%%%%%%%%%%%%%%%%%%%%%%%%%%%%%%%%%%%%%%%%%%%%%%%%%%%%%%%%%%%%%%%%%%%%%%%%%%%%%%%%%%%%%%%%%%%%%%%%%%%%%%%%%%
%%%%%%%%%%%%%%%%%%%%%%%%%%% begin Lemma 2 %%%%%%%%%%%%%%%%%%%%%%%%%%%%%%%%%%%%%%%%%%%%%%%%%%%%%%%%%%%%%%%%%%%%%%%%%%%%%%%%%%%%%%%%%%%%%%%%%%%%%%%%%%%%%%%%%%%%%%%%%%
%%%%%%%%%%%%%%%%%%%%%%%%%%% begin Lemma 2 %%%%%%%%%%%%%%%%%%%%%%%%%%%%%%%%%%%%%%%%%%%%%%%%%%%%%%%%%%%%%%%%%%%%%%%%%%%%%%%%%%%%%%%%%%%%%%%%%%%%%%%%%%%%%%%%%%%%%%%%%%
{\bf Theorem 3.}
{\it  When using ULA antennas, the PDF and CDF of pointing error are obtained as:}
\begin{align}
	\label{ge4}
	& f_{h_p}(h_p) = \frac{1}{\sqrt{\beta_{ty}\beta_{ry}}} 
	G_0^{\frac{2\beta_{ty}\beta_{ry}}{\beta_{ty}+\beta_{ry}} }
	h_p^{\frac{\beta_{ty}+\beta_{ry}}{2\beta_{ty}\beta_{ry}} -1 } \nonumber \\
	& ~~~~\times I_0\left( \frac{\beta_{ty}-\beta_{ry}}{2\beta_{ty}\beta_{ry}}\ln\left(\frac{h_p}{G_0}\right) \right)
\end{align}
\begin{align}
	\label{ge5}
	&F_{h_p}(h_p) = 1 -
	Q\left( \Re_1 \sqrt{\frac{-2\ln\left(\frac{h_p}{G_0}\right)}{\beta_{ty}+\beta_{ry}}}  ,  \Re_2 \sqrt{\frac{-2\ln\left(\frac{h_p}{G_0}\right)}{\beta_{ty}+\beta_{ry}}}  \right) \nonumber \\
	&+  Q\left( \Re_2 \sqrt{\frac{-2\ln\left(\frac{h_p}{G_0}\right)}{\beta_{ty}+\beta_{ry}}}  ,  \Re_1 \sqrt{\frac{-2\ln\left(\frac{h_p}{G_0}\right)}{\beta_{ty}+\beta_{ry}}}  \right)
\end{align}
{\it where $Q(a,b)$ is the Marcum \it Q-function \cite{Marcume_q}, and}
\begin{align}
	\left\{ \!\!\!\!\! \! \!
	\begin{array}{rl}
		&\Re_1 = \frac{\sqrt{1-T_q^4}}{2T_q} \sqrt{\frac{1+T_q}{1-T_q}},\\
		&\Re_2 = \Re_1 \frac{1-T_q}{1+T_q},   ~~~ 
		T_q  = \frac{\sigma_{\theta_\text{min}}}{\sigma_{\theta_\text{max}}}, \\
		&\sigma_{\theta_\text{max}} = \text{max}\{\sigma_{\theta_{q_x}},\sigma_{\theta_{q_y}}\},~~~~
		\sigma_{\theta_\text{min}} = \text{min}\{\sigma_{\theta_{q_x}},\sigma_{\theta_{q_y}}\}
	\end{array} \right. \nonumber
\end{align}
%%%%%%%%%%%%%%%%%%%%%%%%%%% begin PROOF j_u J_u m_u M_u N_{uqw} \mu_{uqx} \mu_{uqy} \sigma_{uqx}  \sigma_{uqy} %%%%%%%%%%%%%%%%%%%%%%%%%%%%%%%%%%%%%%%%%%%%%%%%%%%%%%%%%%%%%%%%%%%%%%%%%%%%%%%%%%%%%%%%%%%%%%%%%%%%%%%%%%%%%%%%%%%%%%%%%%
%%%%%%%%%%%%%%%%%%%%%%%%%%% begin PROOF %%%%%%%%%%%%%%%%%%%%%%%%%%%%%%%%%%%%%%%%%%%%%%%%%%%%%%%%%%%%%%%%%%%%%%%%%%%%%%%%%%%%%%%%%%%%%%%%%%%%%%%%%%%%%%%%%%%%%%%%%%
\begin{IEEEproof}
	Please refer to Appendix \ref{AppC}. 
\end{IEEEproof}

As can be seen, the modified Bessel function of the first kind and the Gamma function have been used in Eqs. \eqref{ge4} and \eqref{ge5}. In order to reduce the complexity of calculations, an approximate model is derived below, which uses only a series of simple multiplication and addition operators to model the PDF and CDF of pointing errors.

%%%%%%%%%%%%%%%%%%%%%%%%%%% begin Lemma 2 %%%%%%%%%%%%%%%%%%%%%%%%%%%%%%%%%%%%%%%%%%%%%%%%%%%%%%%%%%%%%%%%%%%%%%%%%%%%%%%%%%%%%%%%%%%%%%%%%%%%%%%%%%%%%%%%%%%%%%%%%%
%%%%%%%%%%%%%%%%%%%%%%%%%%% begin Lemma 2 %%%%%%%%%%%%%%%%%%%%%%%%%%%%%%%%%%%%%%%%%%%%%%%%%%%%%%%%%%%%%%%%%%%%%%%%%%%%%%%%%%%%%%%%%%%%%%%%%%%%%%%%%%%%%%%%%%%%%%%%%%
%%%%%%%%%%%%%%%%%%%%%%%%%%% begin Lemma 2 %%%%%%%%%%%%%%%%%%%%%%%%%%%%%%%%%%%%%%%%%%%%%%%%%%%%%%%%%%%%%%%%%%%%%%%%%%%%%%%%%%%%%%%%%%%%%%%%%%%%%%%%%%%%%%%%%%%%%%%%%%
{\bf Lemma 1.}
{\it  For linear array antennas, the PDF and CDF of pointing error can be well approximated as:}
\begin{align}
	\label{pe1}
	f_{h_p}(h_p) \simeq  \sum_{n=1}^N \frac{\Re_4 \Re_5}{\Re_3(n)(\beta_{ty}+\beta_{ry})G_0}  
	\left(\frac{h_p}{G_0}\right)^{  \frac{\Re_4}{\beta_{ty}+\beta_{ry}} -1 }
\end{align}
\begin{align}
	\label{ge6}
	F_{h_p}(h_p) \simeq  \sum_{n=1}^N \frac{\Re_5}{\Re_3(n)}
	\left(\frac{h_p}{G_0}\right)^{  \frac{\Re_4}{\beta_{ty}+\beta_{ry}}  }
\end{align}
{\it where}
\begin{align}
	\left\{ \!\!\!\!\! \! \!
	\begin{array}{rl}
		&\Re_3(n) = 1 + \frac{1-T_q^2}{1+T_q^2}\cos\left( \pi \frac{2n-1}{N} \right),\\
		&\Re_4 = \frac{(1+T_q^2)^2}{2T_q^2} \Re_3(n),~~~ \Re_5 = \frac{2T_q}{N(1+T_q^2)}.
	\end{array} \right. \nonumber
\end{align}
%%%%%%%%%%%%%%%%%%%%%%%%%%% begin PROOF j_u J_u m_u M_u N_{uqw} \mu_{uqx} \mu_{uqy} \sigma_{uqx}  \sigma_{uqy} %%%%%%%%%%%%%%%%%%%%%%%%%%%%%%%%%%%%%%%%%%%%%%%%%%%%%%%%%%%%%%%%%%%%%%%%%%%%%%%%%%%%%%%%%%%%%%%%%%%%%%%%%%%%%%%%%%%%%%%%%%
%%%%%%%%%%%%%%%%%%%%%%%%%%% begin PROOF %%%%%%%%%%%%%%%%%%%%%%%%%%%%%%%%%%%%%%%%%%%%%%%%%%%%%%%%%%%%%%%%%%%%%%%%%%%%%%%%%%%%%%%%%%%%%%%%%%%%%%%%%%%%%%%%%%%%%%%%%%
\begin{IEEEproof}
	Please refer to Appendix \ref{AppC}. 
\end{IEEEproof}

\begin{figure}
	\centering
	\subfloat[] {\includegraphics[width=3.0 in]{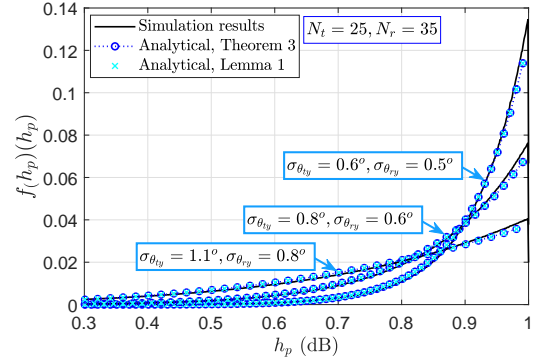}
		\label{bt1}
	}
	\hfill
	\subfloat[] {\includegraphics[width=3.0 in]{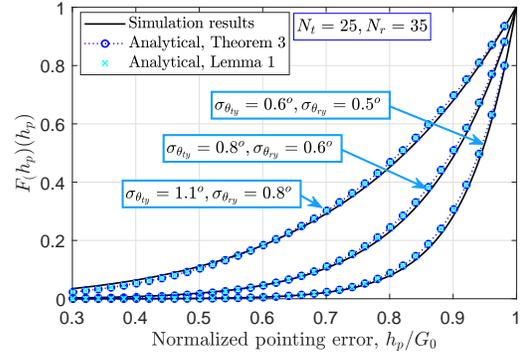}
		\label{bt2}
	}
	\caption{Comparison of the accuracy of the provided analytical results in Theorem 3 and Lemma 1 for ULA antenna with Monte-Carlo simulations for three different states of the intensity of transceiver vibrations (a) the PDF, and (b) the CDF.}
	\label{bt3}
\end{figure}

The closed-form expressions provided in Lemma 1 are very tractable and simple. In addition, the simulation results of Fig. \ref{bt3} verify the accuracy of the analytical expressions of Theorem 3 and Lemma 1 under different values $\sigma_{\theta_{qw}}$s.

%%%%%%%%%%%%%%%%%%%%%%%%%%%%%%%%%%%%%%%%%%%%%%%%%%%%%%%%%%%%%%%%
%%%%%%%%%%%%%%%%%%%%%%%%%%%%%%%%%%%%%%%%%%%%%%%%%%%%%%%%%%%%%%%% VERSUS W_Z
%%%%%%%%%%%%%%%%%%%%%%%%%%%%%%%%%%%%%%%%%%%%%%%%%%%%%%%%%%%%%%%%
%%%%%%%%%%%%%%%%%%%%%%%%%%%%%%%%%%%%%%%%%%%%%%%%%%%%%%%%%%%%%%%%
In some scenarios, we may have a communication link between a stable node and an unstable node. For example, for a ground-to-UAV or a UAV-to-ground station, the intensity of vibrations of the ground node is negligible compared to the UAV node. In this case, the pointing error model provided in Theorem 3 and Lemma 1 can be used with a series of minor changes that are described in {\bf Remark 1}.

{\bf Remark 1.} {\it By following the method used in Appendices A and C, it can be easily shown that the analytical expressions in Theorem 3 and Lemma 1 can be used for a ground-to-UAV link by substituting $\beta_{rx}$ instead of $\beta_{ty}$. Also,  by substituting $\beta_{tx}$ instead of $\beta_{ry}$, we can use Theorem 3 and Lemma 1 for modeling the pointing error of a UAV-to-ground link.}

%---------------------
%---------------------
\section{End-to-End channel Modeling and Performance Analysis}
%---------------------
%---------------------
%For our simulations, we consider standard values for system parameters, as follows. The carrier frequency 
The channel path loss, due to free-space propagation and the molecular absorption loss, is a deterministic parameter and does not change the distribution function of the considered channel. However, in addition to the random parameter due to the pointing error, in some high-frequency typologies, we may also have the effect of small-scale fading. 
Small-scale fading can be caused by various factors such as atmospheric turbulence in rainy weather or multi-path fading for ground links or close to the ground. In this work, we use the $\alpha-\mu$ distribution to model the small-scale fading, which is the most common model for modeling the small-scale fading of high frequency channels.
In the sequel, for the pointing error models provided in Theorems 1-3, we derive the end-to-end channel distribution function of the considered communication link in Theorems 4-6 by considering $\alpha-\mu$ distribution for small-scale fading.

%%%%%%%%%%%%%%%%%%%%%%%%%%%%%%%%%%%%%%%%%%%%%%%%%%%%%%%%%%%%%%%%
%%%%%%%%%%%%%%%%%%%%%%%%%%%%%%%%%%%%%%%%%%%%%%%%%%%%%%%%%%%%%%%% VERSUS W_Z
\begin{figure*}
	\centering
	\subfloat[] {\includegraphics[width=2.3 in]{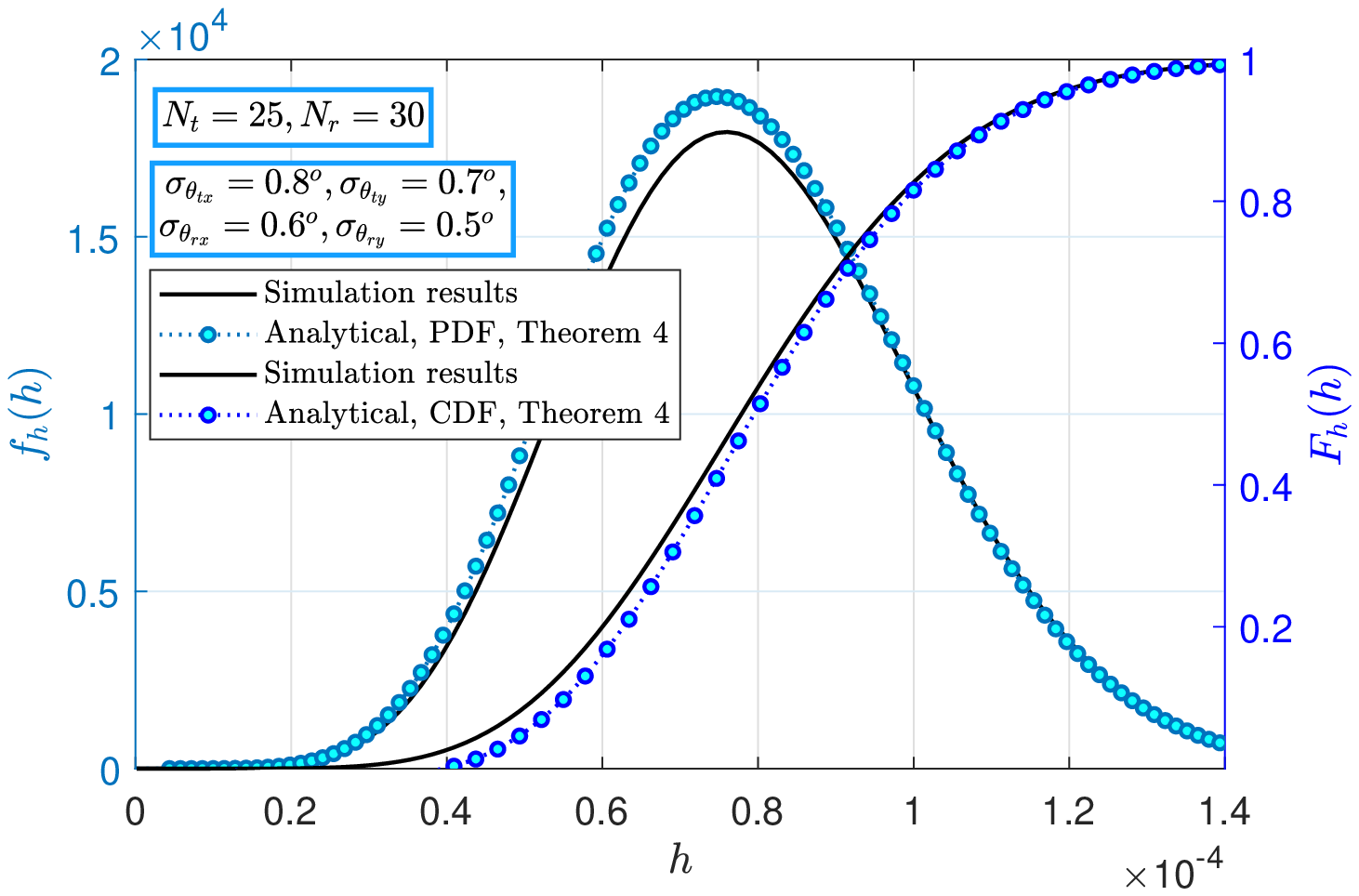}
		\label{br1}
	}
	\hfill
	\subfloat[] {\includegraphics[width=2.3 in]{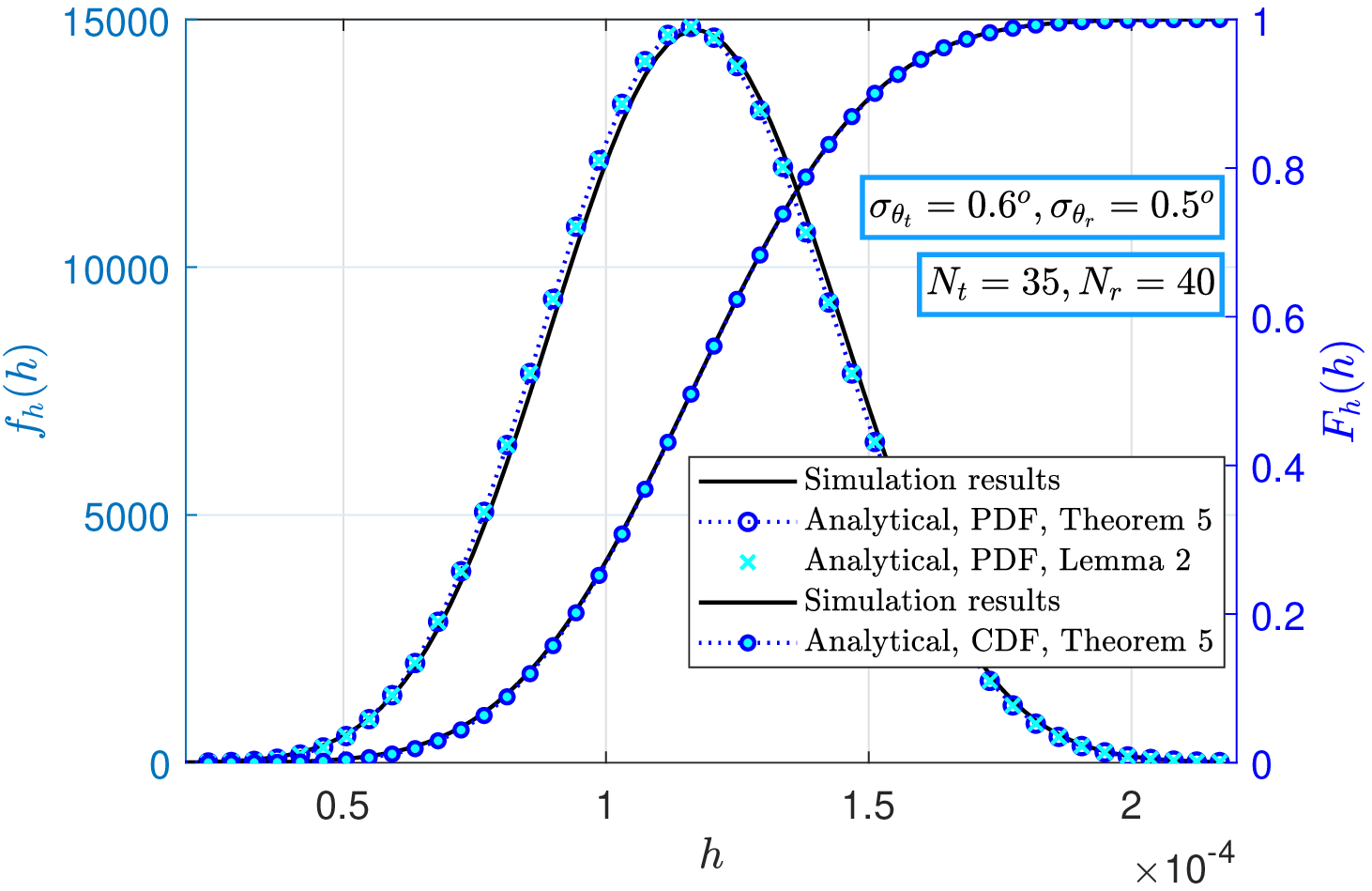}
		\label{br2}
	}
    \hfill
    \subfloat[] {\includegraphics[width=2.3 in]{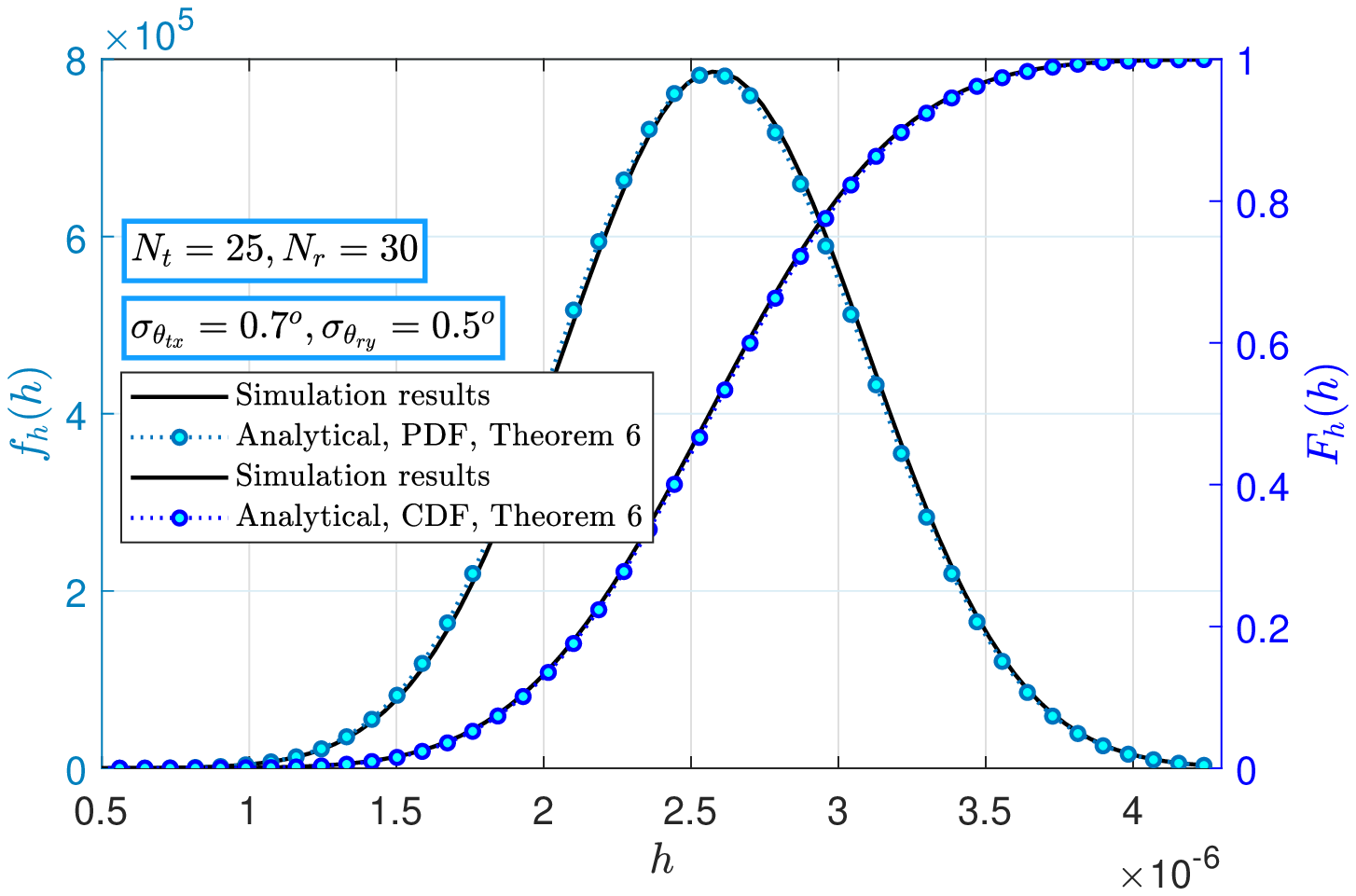}
	\label{br3}
    }
	\caption{Comparison of the accuracy of the provided analytical results with Monte-Carlo simulations: (a) for the general scenario provided in Theorem 4, (b) for a special case provided in Theorem 5 and Lemma 2, and (c) for ULA antenna provided in Theorem 6.}
	\label{br4}
\end{figure*}
%%%%%%%%%%%%%%%%%%%%%%%%%%%%%%%%%%%%%%%%%%%%%%%%%%%%%%%%%%%%%%%%
%%%%%%%%%%%%%%%%%%%%%%%%%%%%%%%%%%%%%%%%%%%%%%%%%%%%%%%%%%%%%%%%

%---------- PDF and CDF of \theta_{tr} for SYMETRC x and y and ASYMETIC t and r

%%%%%%%%%%%%%%%%%%%%%%%%%%% begin Lemma 2 %%%%%%%%%%%%%%%%%%%%%%%%%%%%%%%%%%%%%%%%%%%%%%%%%%%%%%%%%%%%%%%%%%%%%%%%%%%%%%%%%%%%%%%%%%%%%%%%%%%%%%%%%%%%%%%%%%%%%%%%%%
%%%%%%%%%%%%%%%%%%%%%%%%%%% begin Lemma 2 %%%%%%%%%%%%%%%%%%%%%%%%%%%%%%%%%%%%%%%%%%%%%%%%%%%%%%%%%%%%%%%%%%%%%%%%%%%%%%%%%%%%%%%%%%%%%%%%%%%%%%%%%%%%%%%%%%%%%%%%%%
%%%%%%%%%%%%%%%%%%%%%%%%%%% begin Lemma 2 %%%%%%%%%%%%%%%%%%%%%%%%%%%%%%%%%%%%%%%%%%%%%%%%%%%%%%%%%%%%%%%%%%%%%%%%%%%%%%%%%%%%%%%%%%%%%%%%%%%%%%%%%%%%%%%%%%%%%%%%%%
{\bf Theorem 4.}
{\it  Under the general condition when the parameters $\sigma_{\theta_{qw}}$s and $w_{Bq}$s have different values, the PDF of $h$ is obtained as:}
\begin{align}
	\label{b2}
	&f_h(h) \simeq  \frac{A_1  C_g}{2G_0^{\alpha \mu} h_L}   \left( \frac{h}{h_L} \right)^{\alpha \mu -1} 
	\sum_{k=0}^K \frac{\Delta_k   }  {\Gamma(k+2)\beta_1^{k+2}}     e^{-\frac{\mathbb{A}_1(h)}{\alpha^2}} \nonumber\\
	&	\times \frac{1}{\mathbb{A}_1(h)^{k/2+1}} 
	\times\left[ 
	\Gamma\left(\frac{k+2}{2}\right)~\!\! _1F_1\left(\frac{k+2}{2},\frac12,\frac{\mathbb{A}_2(h)}{4} \right) 
	+ \right. \nonumber \\ &\left.
	\frac{\alpha A_3-\mathbb{A}_1(h)}{\alpha\sqrt{\mathbb{A}_1(h)}}
	\Gamma\left(\frac{k+3}{2} \right)~\!\! _1F_1\left(\frac{k+3}{2};\frac32;\frac{\mathbb{A}_2(h)}{4} \right) \right]  
\end{align}
{\it where}
\begin{align}
	\left\{ \!\!\!\!\! \! \!
	\begin{array}{rl}
		&\mathbb{A}_0(h) = \left( \frac{h}{G_0 h_L} \right)^\alpha,~~~
		\mathbb{A}_1(h) = \frac{A_2 \alpha^2}{2} \mathbb{A}_0(h), \\
		&\mathbb{A}_2(h) = \frac{\left(A_3-\mathbb{A}_1(h)/\alpha  \right)^2}{\mathbb{A}_1(h)},\\
		&A_1 = \frac{\alpha \mu^\mu}{\hat{h_a}^{\alpha \mu} \Gamma(\mu)}, ~~~ 
		A_2 =  \frac{ \mu}{\hat{h_a}^\alpha},~~~
		A_3 = \alpha \mu-1/\beta_1
	\end{array} \right. \nonumber
\end{align}
{\it Also, the CDF of $h$ is derived as}
\begin{align}
	\label{b3}
	&F_h(h) = 1 - C_g  \sum_{m=0}^{\mu-1} \sum_{k=0}^K \frac{A_2^m}{\Gamma(m+1)}  \frac{\Delta_k}{\Gamma(k+2)\beta_1^{k+2}}   
	\mathbb{A}_0(h)^m   \nonumber \\ &\times e^{-A_2 \mathbb{A}_0(h) }
	\sum_{j=0}^{k+1} 2^{j-k-2} \binom{k+1}{j}  \left(m \alpha-\frac{1}{\beta_1}-A_2 \mathbb{A}_0(h)\alpha\right)^{k-j+1}  \nonumber \\
	&\times \left(\frac{\alpha^2 A_2 \mathbb{A}_0(h)}{2}\right)^{\frac{j-3}{2}-k}
	\exp\left( \frac{\left(m \alpha-\frac{1}{\beta_1}-A_2 \mathbb{A}_0(h)\alpha\right)^2}{2 \left(\alpha^2 A_2 \mathbb{A}_0(h)\right)} \right) \nonumber \\
	&\times\Gamma\left( \frac{j+1}{2} , \frac{\left(m \alpha-\frac{1}{\beta_1}-A_2 \mathbb{A}_0(h)\alpha\right)^2}{2 \left(\alpha^2 A_2 \mathbb{A}_0(h)\right)}  \right)
\end{align}
%%%%%%%%%%%%%%%%%%%%%%%%%%% begin PROOF j_u J_u m_u M_u N_{uqw} \mu_{uqx} \mu_{uqy} \sigma_{uqx}  \sigma_{uqy} %%%%%%%%%%%%%%%%%%%%%%%%%%%%%%%%%%%%%%%%%%%%%%%%%%%%%%%%%%%%%%%%%%%%%%%%%%%%%%%%%%%%%%%%%%%%%%%%%%%%%%%%%%%%%%%%%%%%%%%%%%
%%%%%%%%%%%%%%%%%%%%%%%%%%% begin PROOF %%%%%%%%%%%%%%%%%%%%%%%%%%%%%%%%%%%%%%%%%%%%%%%%%%%%%%%%%%%%%%%%%%%%%%%%%%%%%%%%%%%%%%%%%%%%%%%%%%%%%%%%%%%%%%%%%%%%%%%%%%
\begin{IEEEproof}
	Please refer to Appendix \ref{AppD}. 
\end{IEEEproof}

%%%%%%%%%%%%%%%%%%%%%%%%%%% begin Lemma 2 %%%%%%%%%%%%%%%%%%%%%%%%%%%%%%%%%%%%%%%%%%%%%%%%%%%%%%%%%%%%%%%%%%%%%%%%%%%%%%%%%%%%%%%%%%%%%%%%%%%%%%%%%%%%%%%%%%%%%%%%%%
%%%%%%%%%%%%%%%%%%%%%%%%%%% begin Lemma 2 %%%%%%%%%%%%%%%%%%%%%%%%%%%%%%%%%%%%%%%%%%%%%%%%%%%%%%%%%%%%%%%%%%%%%%%%%%%%%%%%%%%%%%%%%%%%%%%%%%%%%%%%%%%%%%%%%%%%%%%%%%
%%%%%%%%%%%%%%%%%%%%%%%%%%% begin Lemma 2 %%%%%%%%%%%%%%%%%%%%%%%%%%%%%%%%%%%%%%%%%%%%%%%%%%%%%%%%%%%%%%%%%%%%%%%%%%%%%%%%%%%%%%%%%%%%%%%%%%%%%%%%%%%%%%%%%%%%%%%%%%
{\bf Theorem 5.}
{\it   When $\beta_{tx}=\beta_{ty}=\beta_t$ and $\beta_{rx}=\beta_{ry}=\beta_r$, the PDF and CDF of $h$ are obtained as:}
\begin{align}
	\label{b4}
	&f_h(h) = \frac{ 1}  {\alpha(\beta_t-\beta_r)}   \frac{1}{A_2 h_L G_0} 
	A_1  \left( \mathbb{A}_0(h) \right)^{\mu -\frac{1}{\alpha}-1} \nonumber \\
	& \times \bigg[  (A_2 \mathbb{A}_0(h))^{\frac{V_1}{2}} e^{-\frac{A_2 \mathbb{A}_0(h)}{2}}  \mathbb{W}_{\frac{-V_1}{2},\frac{1-V_1}{2}}(A_2 \mathbb{A}_0(h)) \nonumber \\
	&-(A_2 \mathbb{A}_0(h))^{\frac{V_2}{2}} e^{-\frac{A_2 \mathbb{A}_0(h)}{2}}  \mathbb{W}_{\frac{-V_2}{2},\frac{1-V_2}{2}}(A_2 \mathbb{A}_0(h)) \bigg]
\end{align}
{\it and}
\begin{align}
	\label{b5}
	F_h(h) &= 1 -   \frac{ 1}{\alpha(\beta_t-\beta_r)} \sum_{k=0}^{\mu-1} \frac{\left( A_2\mathbb{A}_0(h) \right)^{k-1}}{\Gamma(k+1)} 
	e^{-\frac{A_2 \mathbb{A}_0(h)}{2} }\\  
	& ~~~\times \bigg[ \left(A_2 \mathbb{A}_0(h)\right)^{\frac{V_3}{2}}  
	\mathbb{W}_{\frac{-V_3}{2},\frac{1-V_3}{2}}\left(A_2 \mathbb{A}_0(h)\right)   \nonumber \\
	&~~~-\left(A_2 \mathbb{A}_0(h)\right)^{\frac{V_4}{2}}  
	\mathbb{W}_{\frac{-V_4}{2},\frac{1-V_4}{2}}\left(A_2 \mathbb{A}_0(h)\right)    \bigg]
\end{align}
{\it where $\mathbb{W}_{-\frac{\nu}{2},\frac{1-\nu}{2}}(u)$ is the Whittaker function \cite{jeffrey2007table}, 
	$V_1 = (1/\alpha\beta_t-\mu+1)$, $V_2 = (1/\alpha\beta_r-\mu+1)$,
	$V_3 = (1/\alpha\beta_t-k+1)$, and $V_4 = (1/\alpha\beta_r-k+1)$.}
%%%%%%%%%%%%%%%%%%%%%%%%%%% begin PROOF j_u J_u m_u M_u N_{uqw} \mu_{uqx} \mu_{uqy} \sigma_{uqx}  \sigma_{uqy} %%%%%%%%%%%%%%%%%%%%%%%%%%%%%%%%%%%%%%%%%%%%%%%%%%%%%%%%%%%%%%%%%%%%%%%%%%%%%%%%%%%%%%%%%%%%%%%%%%%%%%%%%%%%%%%%%%%%%%%%%%
%%%%%%%%%%%%%%%%%%%%%%%%%%% begin PROOF %%%%%%%%%%%%%%%%%%%%%%%%%%%%%%%%%%%%%%%%%%%%%%%%%%%%%%%%%%%%%%%%%%%%%%%%%%%%%%%%%%%%%%%%%%%%%%%%%%%%%%%%%%%%%%%%%%%%%%%%%%
\begin{IEEEproof}
	Please refer to Appendix \ref{AppE}. 
\end{IEEEproof}

%%%%%%%%%%%%%%%%%%%%%%%%%%% begin Lemma 2 %%%%%%%%%%%%%%%%%%%%%%%%%%%%%%%%%%%%%%%%%%%%%%%%%%%%%%%%%%%%%%%%%%%%%%%%%%%%%%%%%%%%%%%%%%%%%%%%%%%%%%%%%%%%%%%%%%%%%%%%%%
%%%%%%%%%%%%%%%%%%%%%%%%%%% begin Lemma 2 %%%%%%%%%%%%%%%%%%%%%%%%%%%%%%%%%%%%%%%%%%%%%%%%%%%%%%%%%%%%%%%%%%%%%%%%%%%%%%%%%%%%%%%%%%%%%%%%%%%%%%%%%%%%%%%%%%%%%%%%%%
%%%%%%%%%%%%%%%%%%%%%%%%%%% begin Lemma 2 %%%%%%%%%%%%%%%%%%%%%%%%%%%%%%%%%%%%%%%%%%%%%%%%%%%%%%%%%%%%%%%%%%%%%%%%%%%%%%%%%%%%%%%%%%%%%%%%%%%%%%%%%%%%%%%%%%%%%%%%%%
{\bf Lemma 2.}
{\it   When $\beta_{tx}=\beta_{ty}=\beta_t$, and $\beta_{rx}=\beta_{ry}=\beta_r$, and under the condition that $\big(\alpha\mu \min\{\beta_t,\beta_r\}\big)>1$, the PDF of $h$ can be approximated as:}
\begin{align}
	\label{b6}
	f_h(h) &= \frac{ 1}  {\alpha(\beta_t-\beta_r)}   \frac{A_1  \left( \mathbb{A}_0(h) \right)^{\mu -\frac{1}{\alpha}-1}}{A_2 h_L G_0} 
	\nonumber \\
	&~~~\times
	\Big[  (A_2 \mathbb{A}_0(h))^{1-s_1} \Gamma\left(s_1,A_2 \mathbb{A}_0(h) \right)  \nonumber \\ &~~~-
	(A_2 \mathbb{A}_0(h))^{1-s_2} \Gamma\left(s_2,A_2 \mathbb{A}_0(h) \right)     \Big]
\end{align}
{\it where $\Gamma(s,x)$, is the upper incomplete gamma function \cite{Incomp_gam}, 
	$s_1 = (-1/\alpha\beta_t+  \mu)$, and $s_2=(-1/\alpha\beta_r+  \mu)$.}
%%%%%%%%%%%%%%%%%%%%%%%%%%% begin PROOF j_u J_u m_u M_u N_{uqw} \mu_{uqx} \mu_{uqy} \sigma_{uqx}  \sigma_{uqy} %%%%%%%%%%%%%%%%%%%%%%%%%%%%%%%%%%%%%%%%%%%%%%%%%%%%%%%%%%%%%%%%%%%%%%%%%%%%%%%%%%%%%%%%%%%%%%%%%%%%%%%%%%%%%%%%%%%%%%%%%%
%%%%%%%%%%%%%%%%%%%%%%%%%%% begin PROOF %%%%%%%%%%%%%%%%%%%%%%%%%%%%%%%%%%%%%%%%%%%%%%%%%%%%%%%%%%%%%%%%%%%%%%%%%%%%%%%%%%%%%%%%%%%%%%%%%%%%%%%%%%%%%%%%%%%%%%%%%%
\begin{IEEEproof}
	Please refer to Appendix \ref{AppE}. 
\end{IEEEproof}

The results of Eq. \eqref{b6} are based on the upper incomplete gamma function $\Gamma(s,x)$, which has faster computation than the Whittaker function. 
However, it should be noted that for $\Gamma(s,x)$, the parameter $s$ must be positive. In Eq. \eqref{b6}, both $s_1$ and $s_2$ are function of the parameters $\alpha$ and $\mu$ related to the small-scale distribution, as well as are function of the parameters $\beta_t$ and $\beta_r$. 
The parameters $\beta_t$ and $\beta_r$ characterize the strength of pointing error. For smaller values of $N_q$ and/or smaller values of $\sigma_{\theta_{qw}}$, the parameters $\beta_t$ and $\beta_r$ decrease, and thus, the values of parameters $s_1$ and $s_2$ becomes negative. As a results, smaller values of $N_q$ and $\sigma_{\theta_{qw}}$, the results of Lemma 2 are not valid.
However, in a wide range of channel parameter values, the results of Lemma 2 are valid.

%%%%%%%%%%%%%%%%%%%%%%%%%%% begin Lemma 2 %%%%%%%%%%%%%%%%%%%%%%%%%%%%%%%%%%%%%%%%%%%%%%%%%%%%%%%%%%%%%%%%%%%%%%%%%%%%%%%%%%%%%%%%%%%%%%%%%%%%%%%%%%%%%%%%%%%%%%%%%%
%%%%%%%%%%%%%%%%%%%%%%%%%%% begin Lemma 2 %%%%%%%%%%%%%%%%%%%%%%%%%%%%%%%%%%%%%%%%%%%%%%%%%%%%%%%%%%%%%%%%%%%%%%%%%%%%%%%%%%%%%%%%%%%%%%%%%%%%%%%%%%%%%%%%%%%%%%%%%%
%%%%%%%%%%%%%%%%%%%%%%%%%%% begin Lemma 2 %%%%%%%%%%%%%%%%%%%%%%%%%%%%%%%%%%%%%%%%%%%%%%%%%%%%%%%%%%%%%%%%%%%%%%%%%%%%%%%%%%%%%%%%%%%%%%%%%%%%%%%%%%%%%%%%%%%%%%%%%%
{\bf Theorem 6.}
{\it  For linear array antenna, the PDF of $h$ is obtained as:}
\begin{align}
	\label{be1}
	& f_h(h) = \sum_{n=1}^N \frac{A_1 \Re_4 \Re_5}{\alpha\Re_3(n)(\beta_{ty}+\beta_{ry})}   \frac{1}{ A_2 h_L G_0}   
	(\mathbb{A}_0(h))^{\mu -1/\alpha-1}   \nonumber \\&~~~~~~~\times
	(A_2 \mathbb{A}_0(h))^{\frac{V_5}{2}} e^{-\frac{A_2 \mathbb{A}_0(h)}{2}}  \mathbb{W}_{\frac{-V_5}{2},\frac{1-V_5}{2}}(A_2 \mathbb{A}_0(h))
\end{align}
{\it and the CDF is calculated as}
\begin{align}
	\label{be2}
	&F_h(h) = 1- \sum_{k=0}^{\mu-1} \sum_{n=1}^N  \frac{1}{\alpha\Gamma(k+1)}  \frac{\Re_4 \Re_5}{\Re_3(n)(\beta_{ty}+\beta_{ry})}  
	\nonumber \\ & 
	\times\left(A_2 \mathbb{A}_0(h)\right)^{k+\frac{V_6}{2}-1} e^{-\frac{A_2 \mathbb{A}_0(h)}{2}}  
	\mathbb{W}_{-\frac{V_6}{2},\frac{1-V_6}{2}}\left(A_2 \mathbb{A}_0(h)\right)
\end{align}
{\it where $V_5 = \frac{\Re_4}{\alpha(\beta_{ty}+\beta_{ry})}-\mu+1)$, and $V_6 = \frac{\Re_4}{\alpha(\beta_{ty}+\beta_{ry})}-k+1)$.}
%%%%%%%%%%%%%%%%%%%%%%%%%%% begin PROOF j_u J_u m_u M_u N_{uqw} \mu_{uqx} \mu_{uqy} \sigma_{uqx}  \sigma_{uqy} %%%%%%%%%%%%%%%%%%%%%%%%%%%%%%%%%%%%%%%%%%%%%%%%%%%%%%%%%%%%%%%%%%%%%%%%%%%%%%%%%%%%%%%%%%%%%%%%%%%%%%%%%%%%%%%%%%%%%%%%%%
%%%%%%%%%%%%%%%%%%%%%%%%%%% begin PROOF %%%%%%%%%%%%%%%%%%%%%%%%%%%%%%%%%%%%%%%%%%%%%%%%%%%%%%%%%%%%%%%%%%%%%%%%%%%%%%%%%%%%%%%%%%%%%%%%%%%%%%%%%%%%%%%%%%%%%%%%%%
\begin{IEEEproof}
	Please refer to Appendix \ref{AppF}. 
\end{IEEEproof}

In Fig. \ref{br4}, the accuracy of the analytical expressions provided in Theorems 4-6 is examined using Monte-Carlo simulations. Here, we set $\alpha=2$, $\mu=4$, $f_c=280$GHz, $\mathcal{K}(f)=2$, and other parameters related to the antenna pattern and pointing errors, as specified in Fig. \ref{br4}.
According to the result obtained from Fig. \ref{br1}, although the analytical results obtained from Theorem 4 are close to the simulation results with an acceptable accuracy, however, a small gap is observed between the simulation results and the results obtained from the approximated analytical expressions in Theorem 2. 
The results of Figs. \ref{br2} and \ref{br3} also show that the analytical results obtained from Theorems 5 and 6 completely match the simulation results.

\begin{figure}
	\begin{center}
		\includegraphics[width=3.35 in]{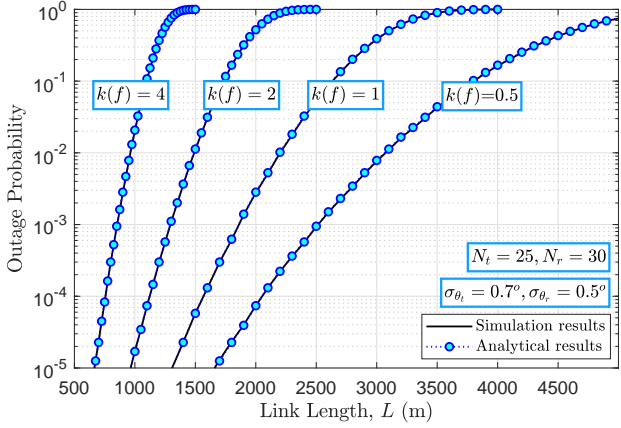}
		\caption{Outage probability of the considered communication link versus link length for different values of molecular absorption loss characterized by $\mathcal{K}(f)$.}
		\label{rn4}
	\end{center}
\end{figure}

\begin{figure}
	\begin{center}
		\includegraphics[width=3.35 in]{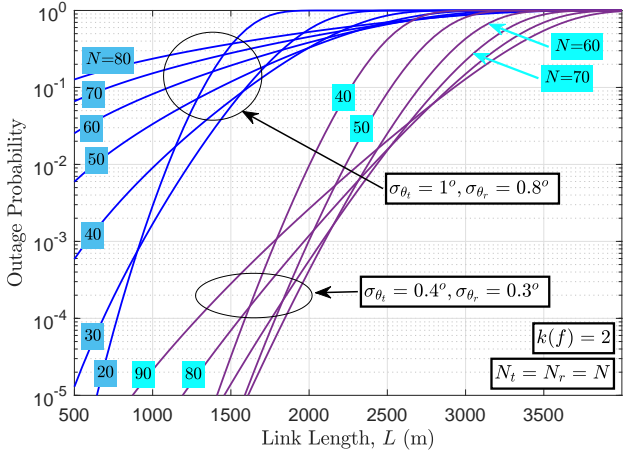}
		\caption{Outage probability of the considered communication link versus link length for two different levels of system instability and different values for $N$.}
		\label{rn5}
	\end{center}
\end{figure}

Finally, in Figs. \ref{rn4} and \ref{rn5}, we evaluate the performance of the considered communication link in term of the outage probability. 
In the simulations, we set transmit power $P_t=10$mW, and SNR threshold $\gamma_{th}=5$dB. Also, for computing thermal noise power, we consider bandwidth $\Delta_f=100$MHz, and receiver's temperature $T=300$K. 
The results of Fig. \ref{rn4} are plotted for different values of molecular absorption loss characterized by $\mathcal{K}(f)=0.5$, 1, 2, and 4.
For example, if we consider that for a quality of service (QoS) with target outage probability lower than $10^{-2}$, it is observed that by increasing $\mathcal{K}(f)$ from 0.5 to 4, the maximum achievable link length to guarantee the requested QoS decreases from 3.1km to lower than 1 km.
There are several ways to increase the maximum achievable link-length of the considered system under the presence of molecular absorption loss. 
However, due to the limited transmission power at higher frequencies, one of the best methods is to increase the gain of the antenna by increasing parameter $N$.
For this aim, the results of Fig. \ref{rn5} are plotted for different values of $N$. 
The results of Fig. \ref{rn5} show an important point that system performance does not necessarily improve with increasing $N$. The reason is that, with increasing antenna gain, the beamwidth of antenna pattern decreases and the system becomes more sensitive to the vibrations of Tx/Rx. To better illustrate this point, the results in Fig. \ref{rn5} are plotted for two different values for $\sigma_{\theta_{qw}}$s. The first category is for $\sigma_{\theta_{t}}=1^o$ and  $\sigma_{\theta_{r}}=0.8^o$. For this state and for target outage probability equal to $10^{-3}$, the optimal value is $N=30$.  
The second category is for $\sigma_{\theta_{t}}=0.4^o$ and  $\sigma_{\theta_{r}}=0.3^o$ that we observe the maximum link length is obtained for $N=60$ in the target outage probability $10^{-3}$.

%----------------------------
%----------------------------
\section{Conclusions}
%----------------------------
%----------------------------
A general pointing error model was investigated for directional mmWave/THz links by taking into account actual antenna patterns. Our derivation is based on the practical assumption that the intensity of instability in Tx and Rx and also the intensity of instability in the Yaw and Pitch directions are different.
%For the considered general scenario, we first derived the PDF and CDF of the pointing error. The case where the Tx and Rx are equipped with uniform linear array antenna, we also derived the PDF and CDF of pointing error. 
%
For lower frequencies as well as higher heights, the small-scale fading effect can be ignored with good accuracy, and as a result, the presented models for pointing errors can be used for end-to-end system analysis. For higher frequencies and in some scenarios, the effect of small-scale fading should be considered.
To this end, using $\alpha-\mu$ distribution, which is a common model for small-scale fading of THz links, the end-to-end PDF and CDF of the considered channel are derived for both planer and linear array antennas. 
The results of this work can be also used to calculate the end-to-end channel distribution in the presence of newer small-scale fading models.
Finally, by employing Monte-Carlo simulations, the accuracy of the analytical expressions was verified and the performance of the system was studied.

%
%%%%%%%%%%%%%%%%%%%%%%%%%%%%%%%%%%%%%%%%%%%%%%%%%%%%%%%%%%%%%%%%
%%%%%%%%%%%%%%%%%%%%%%%%%%%%%%%%%%%%%%%%%%%%%%%%%%%%%%%%%%%%%%%% VERSUS P_T
%%%%%%%%%%%%%%%%%%%%%%%%%%%%%%%%%%%%%%%%%%%%%%%%%%%%%%%%%%%%%%%%
%%%%%%%%%%%%%%%%%%%%%%%%%%%%%%%%%%%%%%%%%%%%%%%%%%%%%%%%%%%%%%%%
%

%
%%%%%%%%%%%%%%%%%%%%%%%%%%%%%%%%%%%%%%%%%%%%%%%%%%%%%%%%%%%%%%%%
%%%%%%%%%%%%%%%%%%%%%%%%%%%%%%%%%%%%%%%%%%%%%%%%%%%%%%%%%%%%%%%% VERSUS P_T
%%%%%%%%%%%%%%%%%%%%%%%%%%%%%%%%%%%%%%%%%%%%%%%%%%%%%%%%%%%%%%%%
%%%%%%%%%%%%%%%%%%%%%%%%%%%%%%%%%%%%%%%%%%%%%%%%%%%%%%%%%%%%%%%%
%

\appendices

%--------------------------------
\section{}
\label{AppA}
The main lobe of the antenna pattern can be well approximated by the Gaussian distribution function as \cite{LetW}
\begin{align}
	\label{b1}
	&G_q(N_q, \theta_{qx},\theta_{qy})          = G_0(N_q) \nonumber \\
	&\times \exp\left( -\frac{\left( \tan^{-1}\left(\sqrt{\tan^2(\theta_{qx})+\tan^2(\theta_{qy})}\right) \right)^2}
	{w^2_B(N_q)}\right), 
\end{align}
where $G_0(N_q)=\pi N_q^2$, and $w_B(N_q)=\frac{1.061}{N_q}$ is the angular beamwidth (called the beam divergence).
For lower values of $\theta_{qx}$ and $\theta_{qy}$, we can approximate RV $\theta_q$ as
\begin{align}
	\label{he7}
	\theta_q \simeq \sqrt{\theta_{qx}^2 + \theta_{qy}^2}.
\end{align}
Substituting Eqs. \eqref{he7} and \eqref{b1} in Eq. \eqref{he5}, we have
\begin{align}
	\label{p1}
	h_p          \simeq G_0(N) \exp\left( -\frac{\theta_{t_x}^2 + \theta_{t_y}^2} {2 w^2_{Bt}}  
	-\frac{\theta_{r_x}^2+\theta_{r_y}^2} {2 w^2_{Br}}\right) 
\end{align}
Let us define parameter $\Theta$ as
\begin{align}
	\label{p2}
	\Theta = \Theta_{t_x} + \Theta_{t_y} + \Theta_{r_x} + \Theta_{r_y}   
\end{align}
where $\Theta_{q_w}=\frac{\theta_{qw}^2}{2 w^2_{Bq}}$. Since $\theta_{qw}\sim\mathcal{N}(0,\sigma_{\theta_{qw}})$, then $\Theta_{q_w}$ follows a gamma distribution as \cite{papoulis2002probability}
\begin{align}
	\label{p3}
	f_{\Theta_{q_w}}(\Theta_{q_w}) = \frac{1}{\sqrt{\beta_{qw} \Theta_{q_w}} \Gamma(1/2)} \exp\left(-\frac{\Theta_{q_w}}{\beta_{qw}}\right)
\end{align}
where $\beta_{qw} = \frac{\sigma_{\theta_{q_w}^2}}{w^2_{Bq}}$.
Let's arrange the parameters $\beta_{qw}$s in ascending order and then define the vector $\bar{\beta}=[\beta_1,...,\beta_4]$, where $\beta_1$ is the least value of $\beta_{q_w}$s and $\beta_4$ is the maximum value of $\beta_{q_w}$s.
Based on the results of \cite[Theorem 1]{moschopoulos1985distribution}, the distribution of RV $\Theta$ is obtained as: 
\begin{align}
	\label{p4}
	f_\Theta(\Theta) = C_g \sum_{k=0}^K \frac{\Delta_k  \Theta^{k+1}\exp\left(-\frac{\Theta}{\beta_1} \right)}
	{\Gamma(k+2)\beta_1^{k+2}}
\end{align}
where
\begin{align}
	\left\{ \!\!\!\!\! \! \!
	\begin{array}{rl}
		&C_g = \prod_{i=1}^4 \sqrt{\beta_1/\beta_i}, \\ 
		&\Delta_k = \frac{1}{k} \sum_{i=1}^k i \gamma_i \Delta_{k+1-i}~~\text{for}~~k=1,...,K, \\
		&\Delta_0=1,~~~
		\gamma_k = \sum_{i=1}^4 \frac{\left(1-{\beta_1}/{\beta_i}\right)^k}{2k}.
	\end{array} \right. \nonumber
\end{align}
From Eq. \eqref{p1}, the distribution of $h_p$ can be obtained as \cite{papoulis2002probability}
\begin{align}
	\label{po1}
	f_{h_p}(h_p) = f_\Theta\left( - \ln\left(\frac{h_p}{G_0}\right) \right) \frac{\text{d}\Theta}{\text{d}h_p}
	= \frac{1}{h_p}   f_\Theta\left( - \ln\left(\frac{h_p}{G_0}\right) \right)
\end{align}
Using Eqs. \eqref{p4} and \eqref{po1}, the closed-form expression for PDF of $h_p$ is derived in Eq. \eqref{gen1}.

From Eq. \eqref{p4}, the CDF of RV $\Theta$ is obtained as
\begin{align}
	\label{p5}
	F_\Theta(\Theta) = C_g \sum_{k=0}^K \frac{\Delta_k } {\Gamma(k+2)\beta_1^{k+2}}
	\int_0^\Theta
	x^{k+1}\exp\left(-\frac{x}{\beta_1} \right)
\end{align}
In the following derivation, we use an integral identity \cite[eq. (2.321.2)]{jeffrey2007table}:
\begin{align}
	\label{p6}
	\int x^n e^{ax} \text{d}x = e^{ax} \left( \sum_{j=0}^n \frac{(-1)^j j! \binom{n}{j}}{a^{j+1}} x^{n-j} \right)
\end{align}  
Using Eq. \eqref{p6}, the CDF of RV $\Theta$ is derived as
\begin{align}
	\label{p7}
	F_\Theta(\Theta) =\Theta_0
	- C_g e^{-\frac{\Theta}{\beta_1}} \sum_{k=0}^K \sum_{j=0}^{k+1} \frac{\Delta_k \Theta^{k+1-j}} {\Gamma(k+2-j)\beta_1^{k+1-j}} 
\end{align}
where
$ \Theta_0 = C_g \sum_{k=0}^K \Delta_k$.
Based on Eq. \eqref{p1}, the CDF of $h_p$ is obtained as
\begin{align}
	\label{p8}
	F_{h_p}(h_p) =\! \text{Prob}\left\{ \Theta > -\! \ln\left(\frac{h_p}{G_0}\right) \right\} = 1\!-\! F_\Theta\left(\! - \!\ln\left(\frac{h_p}{G_0}\right) \right)  
\end{align}
Finally, using Eqs. \eqref{p7} and \eqref{p8}, the closed-form expression for CDF of $h_p$ is derived in Eq. \eqref{gen_cdf_1}.

%%-----------------------------
%%-----------------------------
%%-----------------------------
%%-----------------------------
\section{} \label{AppB}
%%-----------------------------
%%-----------------------------
%%-----------------------------
In this appendix, we obtain the PDF and CDF of $h_p$ for the scenario where the intensity of the vibrations along the Yaw and Pitch axes is the same, approximately.
For this case, using Eqs. \eqref{he5}, \eqref{b1}, and \eqref{he7}, the RV $h_p$ can be formulated as
\begin{align}
	\label{pn1}
	h_p          \simeq G_0(N) \exp\left( -\frac{\theta_{t}^2} {2 w^2_{Bt}}
	-\frac{\theta_r^2} {2 w^2_{Br}}\right).
\end{align}
where $\theta_{q}=\sqrt{\theta_{qx}^2+\theta_{qy}^2}$ and since $\theta_{qw}\sim\mathcal{N}(0,\sigma_{\theta_{qw}})$, the RV $\theta_{qw}$ follows Rayleigh distribution as \cite{papoulis2002probability}
\begin{align}
	\label{pn2}
	f_{\theta_q}(\theta_q) = \frac{\theta_q}{\sigma_{\theta_q}^2}  \exp\left(  -\frac{\theta_q^2}{2 \sigma_{\theta_q}^2}   \right)
\end{align}
Let us define $\theta_{tr}=\frac{\theta_{t}^2} {2 w^2_{Bt}}
+\frac{\theta_r^2} {2 w^2_{Br}}$. The CDF of RV $\theta_{tr}$ conditioned on $\theta_r$ is obtained as:
\begin{align}
	\label{pn3}
	F_{\theta_{tr}|\theta_r}(\theta_{tr}) &= \text{Prob}\left\{ \frac{\theta_{t}^2} {2 w^2_{Bt}}
	+\frac{\theta_r^2} {2 w^2_{Br}} < \theta_{tr}|\theta_r\right\} \nonumber \\          % 
	&= F_t\left( \sqrt{2 w^2_{Bt} \theta_{tr} - \frac{w^2_{Bt}} { w^2_{Br}}\theta_r^2 }\right)
\end{align}
Taking the derivative of Eq. \eqref{pn3} with respect to $\theta_{tr}$, the PDF of RV $\theta_{tr}$ conditioned on RV $\theta_r$ is obtained as
\begin{align}
	\label{pn4}
	f_{\theta_{tr}|\theta_r}(\theta_{tr}) &= \frac{w^2_{Bt}}{\sqrt{2 w^2_{Bt} \theta_{tr} - \frac{w^2_{Bt}} { w^2_{Br}}\theta_r^2 }} \nonumber \\ 
	&~~~\times f_{\theta_t}\left( \sqrt{2 w^2_{Bt} \theta_{tr} - \frac{w^2_{Bt}} { w^2_{Br}}\theta_r^2 } \right)
\end{align}
Using Eqs. \eqref{pn2} and \eqref{pn4}, and after some manipulations, the PDF of $\theta_{tr}$ is derived as
\begin{align}
	\label{pn6}
	f_{\theta_{tr}}(\theta_{tr}) &=      
	\frac{ w^2_{Bt} w^2_{Br} }  {  w^2_{Br} \sigma_{\theta_t}^2   -    w^2_{Bt} \sigma_{\theta_r}^2} \nonumber \\ &\times
	\left[   \exp\left(  -\frac{ w^2_{Bt} }  { \sigma_{\theta_t}^2} \theta_{tr}    \right)  
	-     \exp\left(  -\frac{ w^2_{Br} }  { \sigma_{\theta_r}^2} \theta_{tr}    \right)
	\right]
\end{align}
Now, using Eqs. \eqref{po1} and \eqref{pn6}, and after some derivations, the PDF of pointing error is derived in Eq. \eqref{ge2}.
By integrating Eq. \eqref{pn6}, the CDF of $\theta_{tr}$ is obtained as
\begin{align}
	\label{pn7}
	F_{\theta_{tr}}(\theta_{tr}) &=      
	\frac{ w^2_{Bt} w^2_{Br} }  {  w^2_{Br} \sigma_{\theta_t}^2   -    w^2_{Bt} \sigma_{\theta_r}^2}
	\left[   \frac{\sigma_{\theta_t}^2}{w^2_{Bt}}   \left(1-\exp\left(-\frac{ w^2_{Bt} }  { \sigma_{\theta_t}^2} \theta_{tr}    \right)\right)  
	\right. \nonumber \\   &~~~\left.
	-    \frac{\sigma_{\theta_r}^2}{w^2_{Br}}   \left(1-\exp\left(-\frac{ w^2_{Br} }  { \sigma_{\theta_r}^2} \theta_{tr}    \right)\right)
	%-     \exp\left(  -\frac{ w^2_{Br} }  { \sigma_{\theta_r}^2} \theta_{tr}    \right)
	\right].
\end{align}
Finally, using Eqs. \eqref{p8} and \eqref{pn7}, and after some derivations, the CDF of $h_p$ is derived in Eq. \eqref{ge3}.

%----------------------------
%----------------------------
%----------------------------
\section{}  \label{AppC}
%----------------------------
%----------------------------
%----------------------------
In this section, we derive the PDF and CDF of $h_p$ for the scenario in which the Tx and Rx are equipped with ULA antennas. For ULA antennas, the antenna pattern model provided in Eq. \eqref{f_1} is simplified as \cite{balanis2016antenna}
\begin{align}
	\label{f_11}
	G'_q(N_q,\theta_q, \phi_q) &=
	\left( \frac{\sin\left(\frac{N_q (k d_{y} \sin(\theta_q)\sin(\phi_q))}{2}\right)} 
	{N_q\sin\left(\frac{k d_{y} \sin(\theta_q)\sin(\phi_q)}{2}\right)}\right)^2.
\end{align}
For this scenario, using Eqs. \eqref{he5}, \eqref{b1}, and \eqref{f_11}, the RV $h_p$ is formulated as
\begin{align}
	\label{po5}
	h_p          \simeq G_0 \exp\left( -\frac{\theta_{q_x}^2 + \theta_{q_y}^2} {2 w^2_{Bq}}  \right)
\end{align}
Let us define $\Theta = \frac{\theta_{q_x}^2}{2 w^2_{Bq}}  +  \frac{\theta_{q_y}^2}{2 w^2_{Bq}}$.
Using \cite[eqs. (3) and (5)]{nakagami1960m}, the distribution of $\Theta$ is obtained as
\begin{align}
	\label{po6}
	f_\Theta(\Theta) = \frac{1}{\sqrt{\beta_{qx}\beta_{qy}}} \exp\left(-\frac{\beta_{qx}+\beta_{qy}}{2\beta_{qx}\beta_{qy}}\Theta \right)
	I_0\left( \frac{\beta_{qx}-\beta_{qy}}{2\beta_{qx}\beta_{qy}}\Theta \right)
\end{align}
where $I_0(\cdot)$ is the modified Bessel function of the first kind with order zero.
Using Eqs. \eqref{po1} and \eqref{po6}, the PDF of $h_p$ is derived as
\begin{align}
	\label{po7}
	f_{h_p}(h_p) &= \frac{1}{\sqrt{\beta_{qx}\beta_{qy}}} 
	G_0^{\frac{2\beta_{qx}\beta_{qy}}{\beta_{qx}+\beta_{qy}} }
	h_p^{\frac{\beta_{qx}+\beta_{qy}}{2\beta_{qx}\beta_{qy}} -1 } \nonumber \\ & ~~~\times
	I_0\left( \frac{\beta_{qy}-\beta_{qx}}{2\beta_{qx}\beta_{qy}}\ln\left(\frac{h_p}{G_0}\right) \right)
\end{align}
Based on Eq. \eqref{po7} and using \cite{reference.wolfram_2021_hoytdistribution}, the CDF of $\Theta$ is derived as
\begin{align}
	\label{po8}
	F_\Theta(\Theta) &= Q\left( \Re_1 \sqrt{\frac{2\Theta}{\beta_{qx}+\beta_{qy}}}  ,  \Re_2 \sqrt{\frac{2\Theta}{\beta_{qx}+\beta_{qy}}}  \right)
	\nonumber \\ &~~~
	-  Q\left( \Re_2 \sqrt{\frac{2\Theta}{\beta_{qx}+\beta_{qy}}}  ,   \Re_1 \sqrt{\frac{2\Theta}{\beta_{qx}+\beta_{qy}}}  \right)
\end{align}
where where $Q(a,b)$ is the Marcum {\it Q}-function, and 
\begin{align}
	\left\{ \!\!\!\!\! \! \!
	\begin{array}{rl}
		&\Re_1 = \frac{\sqrt{1-T_q^4}}{2T_q} \sqrt{\frac{1+T_q}{1-T_q}},\\
		&\Re_2 = \Re_1 \frac{1-T_q}{1+T_q},   ~~~ 
		T_q  = \frac{\sigma_{\theta_\text{min}}}{\sigma_{\theta_\text{max}}}, \\
		&\sigma_{\theta_\text{max}} = \text{max}\{\sigma_{\theta_{q_x}},\sigma_{\theta_{q_y}}\},~~~~
		\sigma_{\theta_\text{min}} = \text{min}\{\sigma_{\theta_{q_x}},\sigma_{\theta_{q_y}}\}
	\end{array} \right. \nonumber
\end{align}
Substituting Eq. \eqref{po8} in Eq. \eqref{p8}, the CDF of $h_p$ is obtained as
\begin{align}
	\label{po9}
	F_{h_p}(h_p) &= 1 -
	Q\left( \Re_1 \sqrt{\frac{-2\ln\left(\frac{h_p}{G_0}\right)}{\beta_{qx}+\beta_{qy}}}  ,  \Re_2 \sqrt{\frac{-2\ln\left(\frac{h_p}{G_0}\right)}{\beta_{qx}+\beta_{qy}}}  \right) \nonumber \\ &
	+  Q\left( \Re_2 \sqrt{\frac{-2\ln\left(\frac{h_p}{G_0}\right)}{\beta_{qx}+\beta_{qy}}}  ,  \Re_1 \sqrt{\frac{-2\ln\left(\frac{h_p}{G_0}\right)}{\beta_{qx}+\beta_{qy}}}  \right)
\end{align}

Now in the continuation of this section, we provide simpler analytical expressions for the PDF and CDF of $h_p$.
Using \cite[eq. (5)]{tavares2010efficient}, the CDF of RV $\Theta$ can be well approximated as
\begin{align}
	\label{po10}
	F_\Theta(\Theta) \simeq 1 - \sum_{n=1}^N \frac{\Re_5}{\Re_3(n)}\exp\left(-\frac{\Re_4}{\beta_{qx}+\beta_{qy}} \Theta\right)
\end{align}
where 
\begin{align}
	\left\{ \!\!\!\!\! \! \!
	\begin{array}{rl}
		&\Re_3(n) = 1 + \frac{1-T_q^2}{1+T_q^2}\cos\left( \pi \frac{2n-1}{N} \right),\\
		&\Re_4 = \frac{(1+T_q^2)^2}{2T_q^2} \Re_3(n),~~~ \Re_5 = \frac{2T_q}{N(1+T_q^2)}.
	\end{array} \right. \nonumber
\end{align}
Using Eqs. \eqref{p8} and \eqref{po9}, the CDF of $h_p$ is derived in Eq. \eqref{ge6}. Finally, by taking the derivative of Eq. \eqref{ge6}, the PDF of $h_p$ is derived in Eq. \eqref{pe1}.

%----------------------------------
%----------------------------------
\section{} \label{AppD}
%----------------------------------
%----------------------------------
In this appendix, we obtain the PDF and CDF of $h$ for the considered general scenario.
From Eq. \eqref{he4} and \cite{papoulis2002probability}, we have
\begin{align}
	\label{q1}
	f_h(h) = \int_0^{G_0}  \frac{1}{h_L h_p} f_{h_a}\left( \frac{h}{h_L h_p} \right)  f_{h_p}(h_p) \text{d} h_p
\end{align}
Substituting Eqs. \eqref{he3} and \eqref{gen1} in Eq. \eqref{q1}, we obtain:
\begin{align}
	\label{q2}
	&f_h(h) =  \frac{A_1  C_g}{G_0}  \frac{1}{h_L}   \left( \frac{h}{h_L} \right)^{\alpha \mu -1} 
	\sum_{k=0}^K \frac{\Delta_k   }  {\Gamma(k+2)\beta_1^{k+2}} \times  \nonumber \\ & 
	\int_0^{G_0} \!\!\!  \exp\left(\!\! -A_2 \left( \!\frac{h}{h_L h_p}\! \right)^\alpha \! \right)
	\frac{h_p^{\frac{1}{\beta_1}-\alpha \mu-1}}{G_0^{1/\beta_1-1}}  \left(\! - \!\ln\left(\frac{h_p}{G_0}\right)\! \right)^{k+1} \!\!  \text{d} h_p
\end{align}
where $A_1 = \frac{\alpha \mu^\mu}{\hat{h_a}^{\alpha \mu} \Gamma(\mu)}$, and 
$A_2 =  \frac{ \mu}{\hat{h_a}^\alpha}$.
Applying a change of variable rule $x = - \ln\left(\frac{h_p}{G_0}\right)$, we rewrite Eq. \eqref{q2} as
\begin{align}
	\label{q3}
	&f_h(h) =  \frac{A_1  C_g}{G_0^{\alpha \mu}}  \frac{1}{h_L}   \left( \frac{h}{h_L} \right)^{\alpha \mu -1} 
	\sum_{k=0}^K \frac{\Delta_k   }  {\Gamma(k+2)\beta_1^{k+2}} \nonumber \\ & \times
	\int_0^{\infty}   \exp\left( -A_2 \left( \frac{h}{G_0 h_L} \right)^\alpha e^{\alpha x} \right)
	e^{(\alpha \mu-1/\beta_1)x}  x^{k+1}   \text{d} x
\end{align}
It can be easily shown that for $x>\alpha$, the above integral expression tends to zero.
Therefore, we can approximate $e^{\alpha x}= 1 + \alpha x + \frac{\alpha^2 x^2}{2}$.
Based on this, we approximate Eq. \eqref{q3} as
\begin{align}
	\label{q4}
	&f_h(h) =  \frac{A_1  C_g}{G_0^{\alpha \mu}}  \frac{1}{h_L}   \left( \frac{h}{h_L} \right)^{\alpha \mu -1} 
	\sum_{k=0}^K \frac{\Delta_k   }  {\Gamma(k+2)\beta_1^{k+2}}   \nonumber \\ &\times  \exp\left( -A_2 \left( \frac{h}{G_0 h_L} \right)^\alpha \right)
	\int_0^{\infty}  e^{-A_3 x^2 + A_4 x}     x^{k+1}   \text{d} x
\end{align}
where $A_3 = \left( \frac{A_2}{2} \left( \frac{h}{G_0 h_L} \right)^\alpha  \alpha^2\right)$, and 
$A_4 = \left(\alpha \mu-1/\beta_1-A_2 \left( \frac{h}{G_0 h_L} \right)^\alpha \alpha\right)$.
In the following derivation, we use an integral identity \cite{jeffrey2007table}
\begin{align}
	\label{q5}
	&\int_0^\infty x^n e^{-x^2 + bx} \text{d} x = \frac 12 \left[ 
	b\Gamma\left(\frac n2+1 \right)~\!\! _1F_1\left(\frac n2+1;\frac32;\frac{b^2}{4} \right) \right. \nonumber \\ &~~~~~~~~~~~~~~~~\left.
	+ \Gamma\left(\frac{n+1}{2}\right)~\!\! _1F_1\left(\frac{n+1}{2},\frac12,\frac{b^2}{4} \right) \right]
\end{align}
Using Eqs. \eqref{q4} and \eqref{q5}, and after some manipulations, the closed form expressions for the PDF of $h$ is derived in Eq. \eqref{b2}.

From Eq. \eqref{he4} and \cite{papoulis2002probability}, we have
\begin{align}
	\label{q0}
	F_h(h) = \int_0^{G_0}  F_{h_a}\left( \frac{h}{h_L h_p} \right)  f_{h_p}(h_p) \text{d} h_p
\end{align}
The CDF of $\alpha-\mu$ distribution can be expressed as \cite[eqs. (66) and (67)]{boulogeorgos2019analytical}
\begin{align}
	\label{b0}
	F_{h_a} ( h_a) =
	1 - \sum_{k=0}^{\mu-1} \frac{A_2^k}{\Gamma(k+1)}  h_a^{k \alpha} e^{-A_2 h_a^\alpha}. 
\end{align}
Substituting Eqs. \eqref{gen1} and \eqref{b0} in Eq. \eqref{q0}, we obtain:
\begin{align}
	\label{q6}
	&F_h(h) = 1 -\frac{ C_g}{G_0^{1/\beta_1}}  
	\sum_{m=0}^{\mu-1} \sum_{k=0}^K \frac{A_2^m}{\Gamma(m+1)}  \frac{\Delta_k}{\Gamma(k+2)\beta_1^{k+2}} \times   \nonumber \\ & 
	\int_0^{G_0}    \left( \frac{h}{h_L h_p} \right) ^{m \alpha} e^{-A_2 \left( \frac{h}{h_L h_p} \right)^\alpha}  
	h_p^{\frac{1}{\beta_1}-1}   
	\left(\! - \ln\left(\frac{h_p}{G_0}\right) \right)^{k+1}  \!\!\text{d} h_p
\end{align}
Applying a change of variable rule $x = - \ln\left(\frac{h_p}{G_0}\right)$, Eq. \eqref{q6} is expressed as
\begin{align}
	\label{q7}
	F_h(h) &= 1 - \frac{C_g}{1}  \sum_{m=0}^{\mu-1} \sum_{k=0}^K \frac{A_2^m}{\Gamma(m+1)}  \frac{\Delta_k}{\Gamma(k+2)\beta_1^{k+2}} \nonumber \\ &~~~\times  
	\int_0^{\infty} \left( \frac{h}{h_L G_0} \right) ^{m \alpha} \exp\left(-A_2 \left( \frac{h}{h_L G_0} \right)^\alpha e^{\alpha x} \right) \nonumber \\ &~~~\times
	e^{(m \alpha-\frac{1}{\beta_1}) x}   
	x^{k+1}\text{d} x
\end{align}
In the following derivation, we use an integral identity \cite{jeffrey2007table}
\begin{align}
	\label{q8}
	\int_0^\infty x^n e^{ax^2 + bx} \text{d} x &= \sum_{j=0}^n 2^{j-n-1} \binom{n}{j}  b^{n-j} (-a)^{\frac{j-1}{2}-n} \nonumber \\ &~~~~\times \exp\left(- \frac{b^2}{4 a} \right)
	\Gamma\left( \frac{j+1}{2} ,- \frac{b^2}{4a}  \right)
\end{align}
Finally, using Eqs. \eqref{q7} and \eqref{q8}, and after some manipulations, the closed-form expression for the CDF of $h$ is derived in Eq. \eqref{b3}.

%----------------------------------
%----------------------------------
\section{} \label{AppE}
%----------------------------------
%----------------------------------
In this appendix, we obtain the PDF and CDF of $h$ for the case where the intensity of the vibrations along the Yaw and Pitch axes is the same, approximately.
Substituting Eqs. \eqref{he3} and \eqref{ge2} in Eq. \eqref{q1}, we obtain:
\begin{align}
	\label{n1}
	f_h(h) &= \frac{A_1}  {G_0(\beta_t-\beta_r)}  \int_0^{G_0}  \frac{1}{h_L h_p} \nonumber \\ &~~~\times
	  \left( \frac{h}{h_L h_p} \right)^{\alpha \mu -1}  \exp\left( -A_2 \left( \frac{h}{h_L h_p} \right)^\alpha  \right)     \nonumber \\
	&~~~\times\left(   \left(\frac{h_p}{G_0} \right)^{1/\beta_t-1}   -   \left(\frac{h_p}{G_0} \right)^{1/\beta_r-1}  \right) \text{d} h_p
\end{align}
Applying a change of variable rule $x = \left(\frac{h_p}{G_0}\right)^{-\alpha}$, Eq. \eqref{n1} is rewritten as
\begin{align}
	\label{nn2}
	&f_h(h) = \frac{ 1}  {\alpha(\beta_t-\beta_r)}   \frac{1}{h_L G_0} 
	A_1  \left( \mathbb{A}_0(h) \right)^{\mu -\frac{1}{\alpha}} \times \nonumber \\ &  
	\int_1^{\infty} \exp\left( -A_2 \mathbb{A}_0(h) x \right)     
	\left( x^{-1/\alpha\beta_t+  \mu -1}   -   x^{-1/\alpha\beta_r+  \mu -1}  \right)  \text{d} x
\end{align}
In the following derivation, we use an integral identity \cite{jeffrey2007table}
\begin{align}
	\label{nn3}
	\int_u^\infty  x^{-\nu} {e^{-x}}  \text{d} x = u^{-\nu/2} e^{-u/2}  \mathbb{W}_{-\frac{\nu}{2},\frac{1-\nu}{2}}(u)
\end{align}
Using Eqs. \eqref{nn2} and \eqref{nn3}, the closed-form expressions for the PDF of $h$ is derived in Eq. \eqref{b4}.
Note that the results of Eq. \eqref{b4} are based on the Whittaker function. If $\big(\alpha\mu \min\{\beta_t,\beta_r\}\big)>1$, we can express Eq. \eqref{b4} based on the upper incomplete gamma function $\Gamma(s,x)$, which has faster computation than the Whittaker function. Using Eq. \eqref{nn2} and \cite{Incomp_gam}, and after some derivation, the closed-form expressions for the PDF of $h$ is derived in Eq. \eqref{b6}.

Substituting Eqs. \eqref{b4} and \eqref{b0} in Eq. \eqref{q0}, we obtain:
\begin{align}
	\label{nn4}
	&F_h(h) = 1 -   \frac{ 1}{G_0(\beta_t-\beta_r)} \sum_{k=0}^{\mu-1} \frac{A_2^k}{\Gamma(k+1)}  \int_0^{G_0} \left( \frac{h}{h_L h_p} \right)^{k \alpha}  \times \nonumber \\ &
	\exp\left( -A_2 \left( \frac{h}{h_L h_p} \right)^\alpha \right) \!          
	\left[ \!   \left(\frac{h_p}{G_0} \right)^{1/\beta_t-1}   -   \left(\frac{h_p}{G_0} \right)^{1/\beta_r-1}  \right] \text{d} h_p
\end{align}
Applying a change of variable rule $x = \left(\frac{h_p}{G_0}\right)^{-\alpha}$ and using Eqs. \eqref{nn3} and \eqref{nn4}, the closed-form expressions for the CDF of $h$ is derived in Eq. \eqref{b5}.

%----------------------------------
%----------------------------------
\section{} \label{AppF}
%----------------------------------
%----------------------------------
In this section, we derive the PDF and CDF of $h$ for the case in which the Tx and Rx are equipped with ULA antenna.
Substituting Eqs. \eqref{he3} and \eqref{be1} in Eq. \eqref{q1}, we obtain:
\begin{align}
	\label{pm1}
	f_h(h) &= \sum_{n=1}^N \frac{A_1 \Re_4 \Re_5}{\Re_3(n)(\beta_{ty}+\beta_{ry})G_0}     \int_0^{G_0}  \frac{1}{h_L h_p}  \left( \frac{h}{h_L h_p} \right)^{\alpha \mu -1}  
	\nonumber \\ &\times
	   \exp\left( -A_2 \left( \frac{h}{h_L h_p} \right)^\alpha  \right)       
	\left(\frac{h_p}{G_0}\right)^{  \frac{\Re_4}{\beta_{ty}+\beta_{ry}} -1 } \text{d} h_p
\end{align}
Also, substituting Eqs. \eqref{be1} and \eqref{b0} in Eq. \eqref{q0}, we obtain:
\begin{align}
	\label{pm2}
	&F_h(h) = 1- \sum_{k=0}^{\mu-1} \sum_{n=1}^N  \frac{\Re_4 \Re_5}{\Re_3(n)(\beta_{ty}+\beta_{ry})}  \frac{A_2^k  \left( \frac{h}{G_0 h_L} \right)^{\alpha k}}{\Gamma(k+1)} \times \nonumber \\ & \int_0^{1}  
	\exp\left( -A_2\left( \frac{h}{G_0 h_L} \right)^\alpha y^{-\alpha}  \right)     
	y^{ -k \alpha + \frac{\Re_4}{\beta_{ty}+\beta_{ry}} -1 }  \text{d} y
\end{align}
Using Eqs. \eqref{pm1} and \eqref{pm2} and similar to the method used in Appendix \ref{AppE}, the closed-form expressions for the PDF and CDF of $h$ are derived in Eqs. \eqref{be1} and \eqref{be2}, respectively.

%%%%%%%%%%%%%%%%%%%%%%%%%%%%%%%%%%%%%%%%%%%%%%%%%%%%%%%%%%%%%%
%%%%%%%%%%%%%%%%%%%%%%%%%%%%%%%%%%%%%%%%%%%%%%%%%%%%%%%%%%%%%%
%\bibliographystyle{IEEEtran}
%\balance
%\bibliography{IEEEabrv,myref}


\end{document}